\documentclass[longauth]{aa}

\usepackage{graphicx}
\usepackage{xcolor}
\usepackage{booktabs}
\usepackage{supertabular}
\usepackage[version=4]{mhchem}
\usepackage{ulem}
\usepackage{longtable}
\usepackage{threeparttable}
\usepackage{caption}
\usepackage{sidecap}
\usepackage{float}
\usepackage{placeins}
\usepackage{rotating}

\usepackage{stfloats}
\usepackage{txfonts}
\usepackage{hyperref}
\hypersetup{
    colorlinks=true,
    linkcolor= blue,
    filecolor=blue,      
    urlcolor=blue,
    citecolor= blue,
}%

\begin{document}

   \title{Complex Organic Molecules in Protostars with ALMA Spectral Surveys (COMPASS)}

   \subtitle{IV. Methyl cyanide isotopologues toward BHR71-IRS1}
   \titlerunning{COMPASS. IV. Methyl cyanide isotopologues toward BHR71-IRS1}

   \author{P. Nazari
          \inst{1}
          \and
          A. Coutens\inst{2}
          \and
          J. K. J{\o}rgensen\inst{3}
          \and
          A. Belloche\inst{4}
          \and
          A. L. Plunkett\inst{5}
          \and
          S. Maret\inst{6}
          \and
          D. Harsono\inst{7}
          \and
          M. N. Drozdovskaya\inst{8, 9}
          \and
          J.-E. Lee\inst{10}
          \and
          F. Cruz-S\'aenz de Miera\inst{2}
          \and
          J. Ferrer Asensio\inst{11}
          \and
          M. L. R. van 't Hoff\inst{12}
          \and
          J.-H. Jeong\inst{10}
          \and
          \'A K\'osp\'al\inst{13, 14, 15}
          \and
          B. A. McGuire\inst{5, 16}
          \and
          S. Spezzano\inst{17}
          \and
          Y.-L. Yang\inst{11}
          \and
          H.-S. Yun\inst{10}
          \and
          S. Zeng\inst{11}
          }

   \institute{European Southern Observatory, Karl-Schwarzschild-Strasse 2, 85748 Garching, Germany\\ 
        \email{pooneh.nazari@eso.org}
        \and
        Institut de Recherche en Astrophysique et Plan\'etologie, Universit\'e de Toulouse, CNRS, CNES, 9 av. du Colonel Roche, 31028 Toulouse Cedex 4, France
        \and
        Niels Bohr Institute, University of Copenhagen, Jagtvej 155A, DK-2200 Copenhagen N, Denmark
        \and
        Max-Planck-Institut f{\"u}r Radioastronomie, Auf dem H{\"u}gel 69, 53121 Bonn, Germany
        \and
        National Radio Astronomy Observatory, 520 Edgemont Road, Charlottesville, VA 22903, USA
        \and
        Univ. Grenoble Alpes, CNRS, IPAG, 38000 Grenoble, France
        \and
        Institute of Astronomy, Department of Physics, National Tsinghua University, No. 101, Section 2, Kuang-Fu Road, Hsinchu 30013, Taiwan, R.O.C
        \and
        Physikalisch-Meteorologisches Observatorium Davos und Weltstrahlungszentrum (PMOD/WRC), Dorfstrasse 33, 7260 Davos Dorf, Switzerland
        \and
        Department of Chemistry, Biochemistry and Pharmaceutical Sciences (DCBP), Universit{\"a}t Bern, Freiestrasse 3, 3012 Bern, Switzerland
        \and
        Department of Physics and Astronomy, Seoul National University, 1 Gwanak-ro, Gwanak-gu, Seoul 08826, Republic of Korea
        \and
        Star and Planet Formation Laboratory, RIKEN Pioneering Research Institute, Wako-shi, Saitama, 351-0198, Japan
        \and
        Purdue University, 525 Northwestern Ave, West Lafayette, IN 47907 USA
        \and
        Konkoly Observatory, HUN-REN Research Centre for Astronomy and Earth Sciences, MTA Centre of Excellence, Konkoly-Thege Mikl\'os \'ut 15-17, 1121 Budapest, Hungary
        \and
        Institute of Physics and Astronomy, ELTE E\"otv\"os Lor\'and University, P\'azm\'any P\'eter s\'et\'any 1/A, 1117 Budapest, Hungary
        \and
        Max-Planck-Insitut f\"ur Astronomie, K\"onigstuhl 17, 69117 Heidelberg, Germany
        \and
        Department of Chemistry, Massachusetts Institute of Technology and National Radio Astronomy Observatory, 77 Massachusetts Ave, Cambridge, MA 02139 USA
        \and
        Max-Planck-Institut f{\"u}r extraterrestrische Physik, Giessenbachstrasse 1, 85748 Garching bei M{\"u}nchen, Germany
             }

   \date{Received 22 December 2025; accepted 22 June 2026}

\abstract 
{Methyl cyanide (CH$_3$CN) is one of the most abundant nitrogen-bearing complex organic molecules observed in the gas phase around both low- and high-mass protostars. However, very few observations exist with a broad enough frequency coverage and sufficiently high sensitivity to systematically characterize its minor isotopologues toward such regions.}
{We add to this number by analyzing data toward BHR71-IRS1 taken with the Atacama Large Millimeter/submillimeter Array (ALMA) as part of the Complex Organic Molecules in Protostars with ALMA Spectral Surveys (COMPASS) Large Program over a ${\sim}33$\,GHz frequency range.}
{We studied as many isotopologues of methyl cyanide as possible toward BHR71-IRS1. We found the column densities and excitation temperatures via spectral fitting and assuming local thermodynamic equilibrium. We then compared the derived column density ratios with those toward IRAS 16293-2422 A and B, V883 Ori, G31.41+0.31, and Sgr B2(N2b).}
{We detect CH$_3$CN, CH$_3$CN $v_8$=1, $^{13}$CH$_3$CN, CH$_3^{13}$CN, CH$_3$C$^{15}$N, CH$_2$DCN, and CHD$_2$CN. The column density ratio of $^{13}$CH$_3$CN/CH$_3^{13}$CN is around unity, in agreement with previous observations. The CH$_2$DCN/CH$_3$CN ratios are similar for all the low-mass objects but are ${\sim}1$ order of magnitude higher than those toward Sgr B2(N2b). This could be due to a warmer pre-stellar phase or a different formation timescale of Sgr B2(N2b).}
{The D/H ratios found from methyl cyanide isotopologues agree well with those of methanol toward BHR71-IRS1, likely pointing to the formation and deuteration of methyl cyanide in a similar star-formation phase as methanol (i.e., the pre-stellar phase). We find an increase in the methyl cyanide D/H ratios with the number of deuterium atoms, which suggests that the deuterated isotopologues may form later in the pre-stellar phase. Future COMPASS studies will increase the sample size of low-mass systems with methyl cyanide analysis and potentially examine some of the trends observed here statistically.}


   \maketitle
%
%-------------------------------------------------------------------

\section{Introduction}

Methyl cyanide (CH$_3$CN; also known as acetonitrile) has been observed in the gas phase toward clouds and pre-stellar objects (e.g., \citealt{Blake1987}; \citealt{Vastel2019}; \citealt{Zeng2018}; \citealt{JimenezSerra2021}; \citealt{Megias2023}), protostellar objects (e.g., \citealt{Ewine1995}; \citealt{Jorgensen2005}; \citealt{Beltran2018}; \citealt{Ceccarelli2023}; \citealt{Lee2023}), and protoplanetary disks (e.g., \citealt{Oberg2015}; \citealt{Loomis2018}; \citealt{Ilee2021}). It has also been detected in cometary ices (\citealt{Mumma2011}; \citealt{Rubin2019}) and has most recently been tentatively detected in protostellar ices with the \textit{James Webb} Space Telescope (JWST; \citealt{Nazari2024}). Methyl cyanide is one of the most abundant nitrogen-bearing complex organic molecules (COMs) in the gas around protostars (e.g., \citealt{Calcutt2018}; \citealt{Belloche2020}) and may have a relatively high abundance of ${\sim}1-10$ with respect to OCN$^{-}$ in ices. This suggests that it may be an important nitrogen reservoir in ices (\citealt{Rachid2022}; \citealt{Nazari2025_OCN}) despite being less  abundant than NH$_3$ and NH$_4^+$ (\citealt{Boogert2015}; \citealt{Slavicinska2025}).

Understanding the formation of methyl cyanide is key, as it may be a parent molecule for many nitrogen-bearing COMs (e.g., \citealt{Bulak2021}). However, the formation of methyl cyanide has been a subject of debate for decades (\citealt{Huntress1979}; \citealt{Wilner1994}; \citealt{LeGal2019}). Some studies suggest that it forms in ices and sublimates into gas during the protostellar phase close to the protostar, where the temperatures are higher (\citealt{Walsh2014}; \citealt{Garrod2022}). Potential formation pathways are the hydrogenation of C$_2$N and the reaction of CH$_3$ and CN on the grains (\citealt{Garrod2008}; \citealt{Walsh2014}). Another efficient formation pathway on the grains for CH$_3$CN is through the hydrogenation of H$_2$CCN, where H$_2$CCN first forms in the gas phase (\citealt{Andron2018}). Meanwhile, others find that gas-phase formation routes could also be efficient for this molecule (\citealt{Herbst1980}; \citealt{Calahan2023}; \citealt{Nazari2023CGD}; \citealt{Taniguchi2023}; \citealt{vantHoff2024}). A potential formation pathway in the gas phase is the reaction of HCN with CH$_3^+$ (\citealt{Herbst1980}; \citealt{Garrod2008}). 

Many recent studies have investigated methyl cyanide in the gas phase around low-mass protostars (e.g., \citealt{Bergner2017}; \citealt{Taquet2019}; \citealt{Belloche2020}; \citealt{MartinDomenech2021}; \citealt{Yang2021}; \citealt{Bianchi2022}; \citealt{NAZARI2025_overview}). This molecule has been the focus of many studies because of its special geometry, which provides a unique opportunity to get a handle on the temperature. Particularly, it is a strongly prolate symmetric top molecule (\citealt{Muller2009}). Consequently, its rotational energy states are described by two quantum numbers, J and K. For a given J transition, the K components fall close to each other in the frequency space, producing the so-called K-ladder. Because radiative transitions between the K levels are not allowed, their population is only determined by collisions (\citealt{Yamamoto2017}). This, in addition to the fact that the K transitions cover a range of upper energy levels, makes this molecule ideal for deriving the kinetic temperature. However, previous studies had a sensitivity that was too low or a frequency coverage that was too limited ($\lesssim 5$\,GHz) for the detection of rare isotopologues of this molecule. Detections of $^{13}$C isotopologues are necessary to avoid the optically thick lines of the main isotopologue and thus enable robust measurements of the main isotopologue's  column density. Moreover, detections of its deuterated isotopologues could shed light on how and in what stage of star formation it is formed. Thus, deep observations and broad frequency coverage are crucial for detecting as many minor isotopologues of methyl cyanide as possible with ${\gtrsim}5$ lines spanning a wide range of energy levels to provide accurate measurements of the column densities and excitation temperatures (see the reviews by \citealt{Herbst2009}; \citealt{Jorgensen2020}).

The Atacama Large Millimeter/submillimeter Array (ALMA) has shown a great potential for comprehensive chemical studies through line surveys of protostellar systems. Particularly, a handful of objects have been observed with $> 10$\,GHz frequency coverage with ALMA. These are IRAS 16293-2422 A and B (IRAS 16293A and IRAS 16293B hereafter; \citealt{Jorgensen2016}), Sagittarius (hereafter Sgr) B2(N2) (\citealt{Belloche2025}), G31.41+0.31 (G31.41 hereafter; \citealt{Colzi2021}), and V883 Ori (\citealt{Lee2024}). With this study, carried out as part of the Complex Organic Molecules in Protostars with ALMA Spectral Surveys (COMPASS) Large Program (\citealt{Jorgensen_COMPASS}), we add BHR71-IRS1 to this sample.

BHR71 contains a binary protostellar system consisting of two Class 0 objects, separated by 15.6\arcsec (\citealt{Gavino2024}) and located at a distance of ${\sim}176$\,pc (\citealt{Voirin2018}). This work focuses on IRS1, which has a bolometric luminosity of 9.2\,L$_{\odot}$ (\citealt{Jorgensen_COMPASS}). Complex organics toward BHR71-IRS1 have previously been observed by ALMA (\citealt{Yang2020}). However, in that work, CH$_3$C$^{15}$N was the only methyl cyanide isotopologue detected, and it was only a tentative assignment based on a single transition. In this study we measured column densities of CH$_3$CN isotopologues with a wide frequency coverage (${\sim}33$\,GHz) for the first time toward this object and discuss our values in the context of previous works. BHR71-IRS1 was also part of the Early Planet Formation in Embedded Disks (eDisk; \citealt{Ohashi2023}) ALMA Large Program. \cite{Gavino2024} found that its dust continuum shows a smooth compact profile and placed an upper limit on the central protostellar mass of 0.46\,M$_{\odot}$. Moreover, this object has been observed as a part of the JWST Observations of Young protoStars (JOYS) program (\citealt{ewine2025}). In the infrared it mainly shows gas-phase lines from simple molecules in absorption (\citealt{vanGelder2024}).

This paper is organized as follows. Section \ref{sec:obs_methods} describes our observations and the methods for the spectral fitting. Section \ref{sec:results} presents the column densities measured from the spectral fitting. In Sect. \ref{sec:compare_IRAS_Sgr} we compare our results with previous studies, particularly those on IRAS 16293A and B, V883 Ori, G31.41, and Sgr B2(N2b). Section \ref{sec:like_methanol} discusses the similarity between the deuteration of CH$_3$CN and CH$_3$OH. Finally, Sect. \ref{sec:conclusions} concludes with our findings.

\section{Observations and methods}
\label{sec:obs_methods}
\subsection{Observational data}

\begin{figure}
    \centering
    \includegraphics[width=\columnwidth]{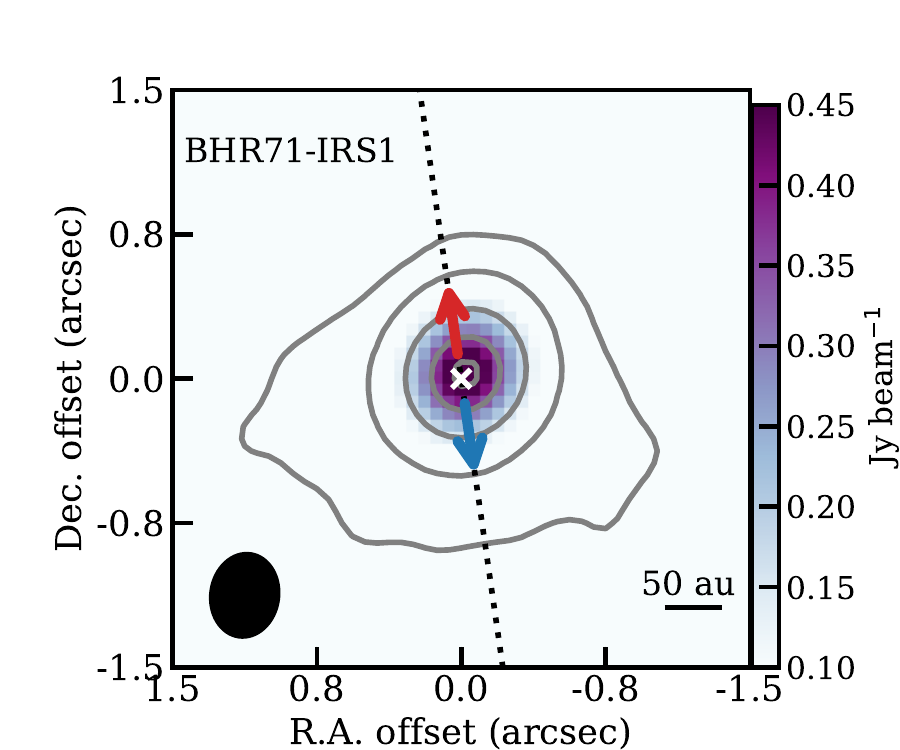}
    \caption{Continuum emission of BHR71-IRS1 at 290.406\,GHz. The contours show the [30, 100, 400, 850, 1200]$\sigma$ levels of the continuum, where $\sigma$ is $4\times10^{-4}$\,Jy\,beam$^{-1}$. The central cross indicates where the spectra were extracted, and the dotted line shows the outflow axis based on \cite{Gavino2024}. The blue and red arrows indicate the direction of blue- and redshifted outflow emission, respectively. The beam is shown in the bottom left.}
    \label{fig:cont}
\end{figure}

The data of BHR71-IRS1 were taken as a part of the COMPASS Large Program (project ID: 2022.1.00316.L\footnote{\href{https://erda.dk/vgrid/COMPASS/}{https://erda.dk/vgrid/COMPASS/}}; \citealt{Jorgensen_COMPASS}). The data reduction is explained in detail in \cite{Plunkett_COMPASS}, and the data reduction pipeline is publicly available on GitHub (\citealt{Maret25}). Here we provide a brief description. The strategy of COMPASS was to perform unbiased spectral surveys of 11 protostellar regions between 279 and 312\,GHz with ALMA. The data were calibrated using CASA (\citealt{McMullin2007}; \citealt{CASA2022}) and imaged using IMAGER\footnote{https://imager.oasu.u-bordeaux.fr}. As part of this program, BHR71-IRS1 was observed with nine tunings; each tuning included four back-to-back spectral windows. BHR71-IRS1 was the only object that was fully observed in Cycle 9. The angular resolution for its observations was $0.42\arcsec\times0.33\arcsec$ to $0.51\arcsec\times0.38\arcsec$ (${\sim}74\times58$\,au to ${\sim}90\times67$\,au), which changed with frequency. The spectral resolution of the data was ${\sim}0.47-0.52$\,km\,s$^{-1}$. The lines of interest have full widths at half maximum (FWHMs) of ${\sim}3$\,km\,s$^{-1}$ toward BHR71-IRS1 (Sect. \ref{sec:spec_fit}) and are thus spectrally resolved. The continuum was subtracted for BHR71-IRS1 in each pixel following \cite{Jorgensen2016}. This was to avoid over-subtraction in the most line-confused regions. The spectrum was then extracted from the continuum peak position analyzed in \citet[$\alpha_{\rm{ICRS}}= 12^{\rm h}01^{\rm m}36.476^{\rm s}$, $\delta_{\rm ICRS} = -65^{\circ}08\arcmin49.37\arcsec$; see Fig. \ref{fig:cont}]{Coutens_COMPASS}. The root mean square (rms) noise for the spectrum varies between ${\sim}0.17$ and ${\sim}0.25$\,K (\citealt{Plunkett_COMPASS}).

\begin{figure*}
    \centering
    \includegraphics[width=19cm]{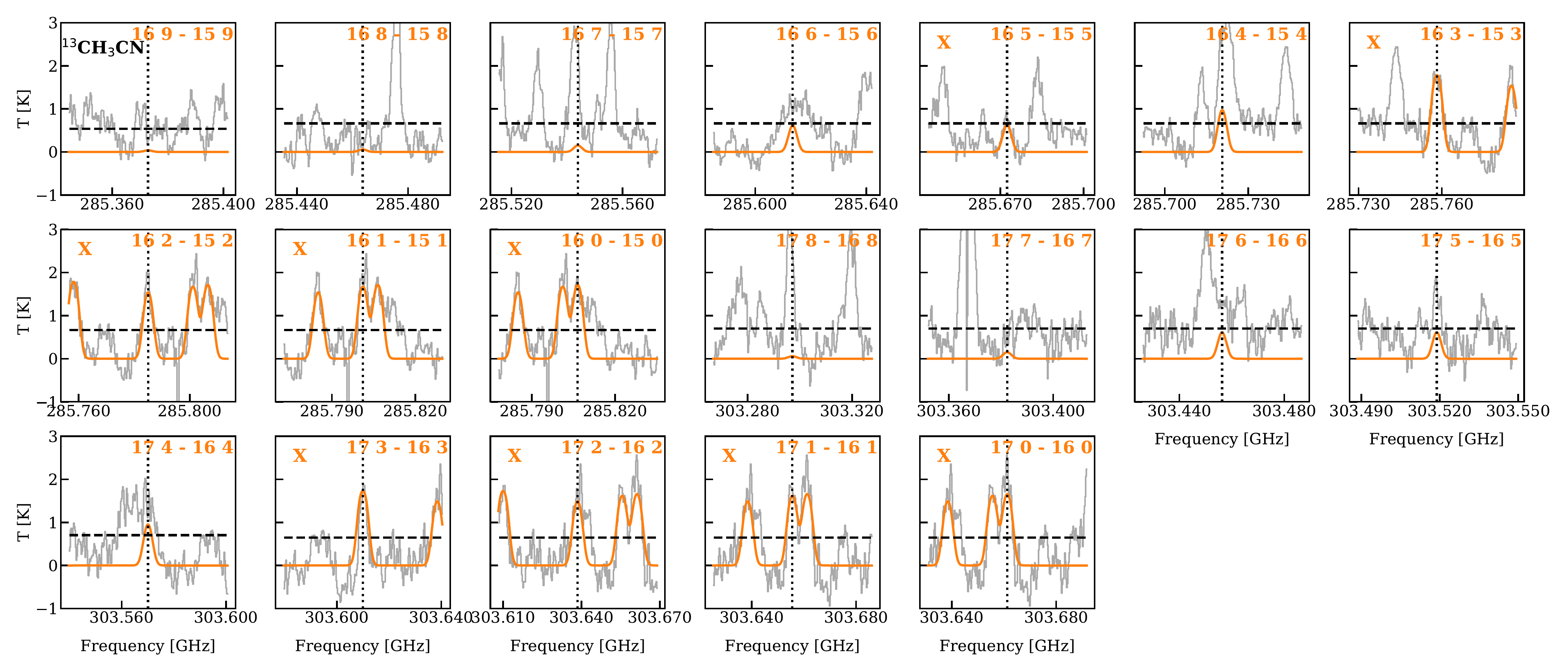}
    \caption{Best-fit model (see Table \ref{tab:results}) for $^{13}$CH$_3$CN (in orange) shown on top of the data (in gray) toward the continuum peak. All lines with $E_{\rm up} < 700$\,K and $A_{\rm ij} >10^{-6}$\,s$^{-1}$ are shown. This limit was chosen to show all the detected lines and a few non-detected ones. The non-detected lines are presented to emphasize that the model does not overestimate any transition. The quantum numbers for the transitions are given in the top-right corner of each panel. The dotted line indicates the rest frequency of each emission line. The horizontal dashed line indicates the $3\sigma$ level. The detected and unblended lines that were fitted are marked with an X in the top-left corner of each panel. The fitting was done using the spectra including the continuum, but for presentation purposes we show the final fits on the continuum-subtracted spectra. }
    \label{fig:13CH3CN}
\end{figure*}

\subsection{Spectroscopic data}
\label{sec:Spectroscopic_data}

The spectral data for each molecule were taken from the Cologne Database for Molecular Spectroscopy (CDMS; \citealt{Muller2001}; \citealt{Muller2005}; \citealt{Endres2016}). Appendix \ref{app:transitions} gives the references for the spectroscopic data and presents the transitions that are covered in the data (Tables \ref{tab:transition_CH3CN}-\ref{tab:transition_CHD2CN}). It is worth noting that most methyl cyanide isotopologues are symmetric tops (\citealt{Muller2016}). This allows the transitions of each molecule to be described by two quantum numbers, J and K. The J quantum number describes the total angular momentum, and the K quantum number describes the projection of the angular momentum along the internal symmetry axis of the molecule. This does not hold for CH$_2$DCN or CHD$_2$CN, which are asymmetric tops (\citealt{LeGuennec1992}) and thus a third quantum number is needed to describe their transitions. Moreover, methyl cyanide in the vibrationally excited state also requires a third quantum number to describe its vibrational state. Vibrational contributions to the partition functions of CH$_3$CN and CH$_3$CN $v_8$=1 are considered for states up to 1200\,cm$^{-1}$ in the database. Higher states contribute less than ${\sim}3\%$ at 300\,K and, thus, are ignored here. However, these vibrational contributions have not been considered for the partition functions of the minor isotopologues in the CDMS database. Thus, as a first-order approximation, we calculated the vibrational correction factors for $^{13}$CH$_3$CN, CH$_3^{13}$CN, CH$_3$C$^{15}$N, CH$_2$DCN, CHD$_2$CN, $^{13}$CH$_3^{13}$CN, and CD$_3$CN as 1.03, 1.12, 1.43 (1.06 at 150\,K), 1.10, 1.09, 1.06, and 1.09 for their respective fitted excitation temperatures.

\subsection{Spectral fitting}
\label{sec:spec_fit}
In this work we focused on identifying the lines of methyl cyanide isotopologues. We fitted the lines of the detected species using the CASSIS\footnote{\href{https://cassis.irap.omp.eu}{https://cassis.irap.omp.eu}, CASSIS has been developed by IRAP-UPS/CNRS.} spectral analysis tool (\citealt{Vastel2015}) to derive the column densities. We assumed local thermodynamic equilibrium (LTE). As shown in Appendix C of \cite{Coutens_COMPASS}, LTE holds well for methanol toward this object in the inner high-density regions. Our considered molecules also trace the inner dense regions of the protostellar system (Fig. \ref{fig:maps}), and for each molecule a single temperature fits all the lines well; therefore, the LTE assumption is most likely valid. The spectrum was fitted as a whole for the unblended lines of each molecule simultaneously. We fitted the lines on top of the continuum by feeding the continuum and spectrum extracted with the continuum into CASSIS. Including the continuum in the fitting procedure ensures that the frequency dependence of the continuum is correctly considered. Moreover, it avoids an underestimation of column densities. The CASSIS formalism and the mathematical expressions behind this procedure are explained in detail in \cite{Coutens_COMPASS}. Figures \ref{fig:13CH3CN} and \ref{fig:CH3C-13-N}-\ref{fig:CH3CN} show our best-fit models for all detected isotopologues of methyl cyanide.

For the spectral fitting, we fixed the systemic velocity ($v_{\rm lsr}$), the line width (FWHM), and the source size (Table \ref{tab:results}). We fixed $v_{\rm lsr}$ to $-4.45$\,km\,s$^{-1}$. This value is the same as that found toward BHR71-IRS1 for methanol (\citealt{Coutens_COMPASS}) and fits the methyl cyanide lines well. We fitted Gaussians to ${\sim}10$ unblended lines of methyl cyanide isotopologues and found that the FWHM falls in the range ${\sim}3.4-3.8$\,km\,s$^{-1}$. Therefore, we fixed the FWHM to 3.6\,km\,s$^{-1}$. The uncertainty on the FWHM is ${\sim}0.5$\,km\,s$^{-1}$, mainly resulting from the variations between unblended lines. Within the uncertainties, the line widths of methyl cyanide could agree with those of methanol toward BHR71-IRS1 (\citealt{Coutens_COMPASS}), but the methanol line widths were generally smaller than what is found here by ${\sim}0.8$\,km\,s$^{-1}$.

We investigated the possibility of finding the methyl cyanide emission size by fitting the optically thin and optically thick lines of CH$_3$CN $v$=0, $^{13}$CH$_3$CN, and CH$_3^{13}$CN simultaneously. The mathematical expressions and physical reasons that allow this measurement are explained in detail in Appendix A of \cite{Coutens_COMPASS}. In simple words, the intensity of a line will increase with column density only if the lines are optically thin. Once a line becomes optically thick, an increase in column density will result in a saturation of the intensity. The intensity of a spatially unresolved optically thick line scales only with the temperature and the emission size. Therefore, if both optically thin and thick lines are present, only certain combinations of column density, source size, and temperature can give a good fit. CASSIS accounts for the line optical depth in the calculation of the synthetic spectra. Thus, for this measurement, we varied the source size between 0.01\arcsec and 0.2\arcsec\ while finding the best possible column density and excitation temperature for each source size. We found source sizes between ${\sim}0.04\arcsec$ and ${\sim}0.13\arcsec$ with best-fit values of 0.07\arcsec, 0.09\arcsec, and 0.05\arcsec for CH$_3$CN $v$=0, $^{13}$CH$_3$CN, and CH$_3^{13}$CN, respectively. This range is in line with the sizes found for methanol and its isotopologues toward the same object (\citealt{Coutens_COMPASS}). However, the number of detected and unblended lines of CH$_3$CN $v$=0, $^{13}$CH$_3$CN, and CH$_3^{13}$CN (Table \ref{tab:results}) compared to those of methanol are much smaller, with most methanol isotopologues having more than 20 optically thin and thick lines detected toward BHR71-IRS1 (\citealt{Coutens_COMPASS}). Therefore, we opted to fix the source size for the column density and temperature measurement of methyl cyanide isotopologues to 0.07\arcsec, which is the same as that used for methanol in \cite{Coutens_COMPASS} and is roughly in the middle of the range found for CH$_3$CN $v$=0, $^{13}$CH$_3$CN, and CH$_3^{13}$CN. This choice is further supported by the similar emitting regions of $^{13}$CH$_3$CN and $^{13}$CH$_3$OH lines with comparable upper energy levels within the limitation of our beam (Appendix \ref{app:fit_size}). Figures \ref{fig:13CH3CN} and \ref{fig:CH3C-13-N}-\ref{fig:CHD2CN} show that the above assumptions result in models that fit the data well. It is worth noting that the $K=0,1,2$ lines of $^{13}$CH$_3$CN, CH$_3^{13}$CN, and CH$_3$C$^{15}$N have similar intensities, which is likely due to the higher optical depth of these lines ($1\lesssim \tau < 2$).

After fixing the aforementioned parameters (i.e., FWHM, V$_{\rm lsr}$, and source size), the column densities and excitation temperatures were fitted. The fitting process was similar to the grid fitting method of \cite{vanGelder2020} and consisted of finding the minimum $\chi^2$ for the fits at grid points with different column density and excitation temperature combinations. In this procedure, only the unblended lines were fitted. The unblended lines were identified by first fitting a preliminary model that included other prominent molecules, and the lines with major contributions from those molecules were excluded when fitting for methyl cyanide. We note that this preliminary model was only used to identify the blended lines, which are eliminated when fitting for methyl cyanide. Thus, possible changes in the final models of the contaminating species will not affect our column densities for methyl cyanide. The fitted unblended methyl cyanide lines are marked in Figs. \ref{fig:13CH3CN} and \ref{fig:CH3C-13-N}-\ref{fig:CHD2CN}. The fitting process included an exploratory phase in which we first found the best-fit model using a sparse but broad set of $N$ values. We then made a narrower but finer array of column densities around those best-fit $N$. Column densities for the smaller arrays were varied over two orders of magnitude with 0.1 step sizes in $\log$ space. For the final grid, we also varied the excitation temperature in 10\,K steps over 50-300\,K or 50-500\,K, depending on the availability of data for the partition functions of the molecules at higher temperatures. Once the best-fit models were identified for this finer grid, we ensured that they were well within the grid boundaries. In the fitting, we considered a 10\% calibration uncertainty and a line rms that varied per spectral window as measured in \cite{Plunkett_COMPASS}. The 2$\sigma$ uncertainties from the grid were calculated on the column densities and excitation temperatures. The uncertainties on the column density ratios were simply calculated by propagating these uncertainties.

\section{Fitting results}
\label{sec:results}

\subsection{Detection statistics}
From the isotopologues of methyl cyanide, CH$_3$CN, $^{13}$CH$_3$CN, CH$_3^{13}$CN, CH$_3$C$^{15}$N, CH$_2$DCN, and CHD$_2$CN were detected. We detected neither $^{13}$CH$_3^{13}$CN nor CD$_3$CN. The spectroscopy for species with combinations of $^{13}$C, $^{15}$N, and D was not available. The upper limits on $^{13}$CH$_3^{13}$CN and CD$_3$CN are higher than the column densities expected from the assumed isotopologue ratios. Therefore, the non-detection of these two molecules is due to a lack of sensitivity. We also detected vibrationally excited methyl cyanide, CH$_3$CN $v_8$=1, but CH$_3$CN $v_8$=2 and vibrationally excited $^{13}$CH$_3$CN or CH$_3^{13}$CN were not detected. For all the detected molecules, we identified at least five unblended lines above the $3\sigma$ level. The number of detected lines for each molecule is presented in Table \ref{tab:results}. The transitions that were not detected are all predicted to have peak intensities below the $3\sigma$ level by our best-fit models or to be blended with emission from other species (Figs. \ref{fig:13CH3CN} and \ref{fig:CH3C-13-N}-\ref{fig:CHD2CN}).

\subsection{Column densities and excitation temperatures}
\label{sec:diff_extract}

\begin{table}
\centering
\renewcommand{\arraystretch}{1.3}
    \caption{Best-fitting parameters at the continuum peak.}
    \label{tab:results}
    \resizebox{0.9\columnwidth}{!}{\begin{tabular}{@{\extracolsep{1mm}}*{4}{l}}
          \toprule
          \toprule      
       Species  & No. of  &  N & T \\ 
         & lines$^{\rm a}$ & [cm$^{-2}$] & [K]\\
        \midrule 

CH$_3$CN $v=0$ & 10 & $^{\rm b}$4.6$^{+0.4}_{-0.5}$ $\times 10^{17}$ & -- \\
 &  & $^{\rm c}$2.5$^{+0.3}_{-0.5}$ $\times 10^{17}$ & [150] \\
CH$_3$CN $v8=1$ & 7 & 3.2$^{+1.8}_{-0.3}$ $\times 10^{17}$ & 140$^{+5}_{-10}$ \\
$^{13}$CH$_3$CN & 9 & 6.5$^{+0.8}_{-1.3}$ $\times 10^{15}$ & 120$^{+20}_{-5}$ \\
CH$_3^{13}$CN & 5 & 7.1$^{+0.9}_{-0.8}$ $\times 10^{15}$ & 180$^{+20}_{-20}$ \\
CH$_3$C$^{15}$N & 8 & 3.6$^{+0.9}_{-0.4}$ $\times 10^{15}$ & 280$^{+130}_{-40}$ \\
 & & 2.7$^{+0.3}_{-0.5}$ $\times 10^{15}$ & [150]\\
CH$_2$DCN & 11 & 1.7$^{+0.2}_{-0.2}$ $\times 10^{16}$ & 170$^{+10}_{-5}$ \\
CHD$_2$CN & 12 & 6.9$^{+0.8}_{-0.7}$ $\times 10^{15}$ & 150$^{+20}_{-10}$ \\
$^{13}$CH$_3^{13}$CN & 0 &$^{\rm d}$<\,4.2 $\times 10^{14}$ &[150] \\
CD$_3$CN & 0 &$^{\rm d}$<\,6.5 $\times 10^{14}$ &[150] \\
\bottomrule
        \end{tabular}}
        \tablefoot{Fitted column density (including vibrational correction factors) and excitation temperature at the continuum peak position. The source size, FWHM, and $v_{\rm LSR}$ are fixed at 0.07\arcsec, 3.6\,km\,s$^{-1}$, and -4.45\,km\,s$^{-1}$. The $2\sigma$ uncertainties from the grid are presented. The brackets indicate when the temperature is fixed to the average of those for $^{13}$CH$_3$CN and CH$_3^{13}$CN. (a) Number of detected and unblended lines of each molecule. (b) Column density measured by multiplying the average column density of $^{13}$CH$_3$CN and CH$_3^{13}$CN by $^{12}$C/$^{13}$C = 68 (\citealt{Milam2005}). This column density was adopted as the reference value for methyl cyanide unless otherwise stated. (c) Column density measured from two lines of methyl cyanide that may be optically thin. Considering the weaker of the two lines alone in the fits increases the column densities by a factor of ${\sim}2-3$. (d) 3$\sigma$ upper limits measured by considering the area below the brightest line.}
        
\end{table}

The measured excitation temperatures for most isotopologues of methyl cyanide agree with each other and vary between ${\sim}120$ and ${\sim}200$\,K. The only molecule that shows a higher excitation temperature (${\sim}280$\,K) is CH$_3$C$^{15}$N. However, the uncertainties on its $T_{\rm ex}$ are also relatively large (${\sim}50-100$\,K). Therefore, we provide the column density of this molecule in Table \ref{tab:results} at 150\,K as well. The temperatures for most methyl cyanide isotopologues agree with those found for most methanol isotopologues toward the same object (${\sim}140-170$\,K; \citealt{Coutens_COMPASS}). This may indicate that the two molecules trace regions with similar temperatures around the protostar and further justifies our choice of using the same emitting size for them in the fitting process (Sect. \ref{sec:spec_fit}).

The fitted column densities from the spectra extracted at the continuum peak are given in Table \ref{tab:results}. The methyl cyanide isotopologue column densities vary between ${\sim}10^{15}$ and ${\sim}5\times10^{17}$\,cm$^{-2}$. We measured the column density of the main isotopologue in three ways: (1) We assumed a $^{12}$C/$^{13}$C ratio of 68 (\citealt{Milam2005}) and found the column density of CH$_3$CN from the average of the $^{13}$CH$_3$CN and CH$_3^{13}$CN column densities. (2) We found the column density of methyl cyanide from its potentially optically thin lines (i.e., the two weakest unblended lines with $\tau {<}1$) while fixing the excitation temperature to the average of those found for $^{13}$CH$_3$CN and CH$_3^{13}$CN (Fig. \ref{fig:CH3CN}). (3) We found the methyl cyanide column density using its vibrationally excited lines (CH$_3$CN $v_8$=1). These three different methods result in column densities that generally agree with one another within a factor of ${\sim}2$. The reason that methods 2 and 3 result in a factor of ${\sim}2$ lower values than that found through method 1 could be the high optical depth of the CH$_3$CN $v$=0 and CH$_3$CN $v_8$=1 lines. For example, if only the weakest line of CH$_3$CN $v$=0 were fitted, the column density would increase by a factor of ${\sim}2-3$. Moreover, almost all the fitted lines of CH$_3$CN $v_8$=1 have optical depths of $1 \lesssim \tau < 2$. Another possibility is that the region that the CH$_3$CN $v_8$=1 lines trace is smaller than the region traced by $^{13}$CH$_3$CN and CH$_3^{13}$CN. This is relevant because all the detected CH$_3$CN $v_8$=1 lines have $E_{\rm up}>500$\,K (Table \ref{tab:transition_CH3CN_v8}). Despite this, our ALMA moment 0 maps (Fig. \ref{fig:maps}) show negligible differences across lines of varying upper energy levels, likely because the bulk emission is unresolved within the beam. A final possibility is that the $^{12}$C/$^{13}$C ratio is lower than 68. However, considering the low number of potentially optically thin lines of CH$_3$CN $v$=0 (i.e., 2) and the high $E_{\rm up}$ of all the lines of CH$_3$CN $v_8$=1 covered in our data, we opted to use the column density found through method 1 as the reference throughout this work, unless otherwise stated.

\section{Comparison with other sources}
\label{sec:compare_IRAS_Sgr}

\begin{figure}
  \resizebox{\columnwidth}{!}{\includegraphics{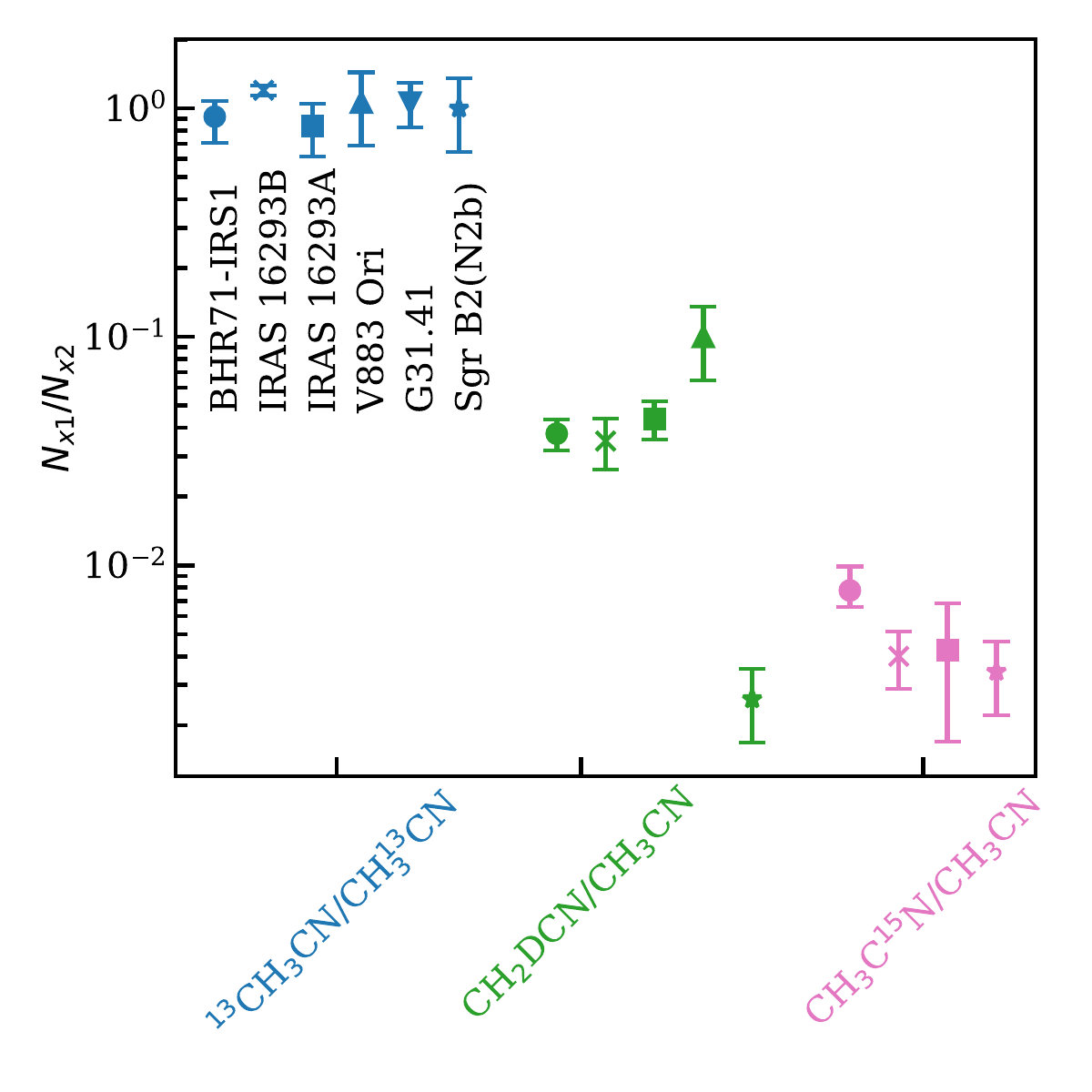}}
  \caption{Methyl cyanide isotopologues' column density ratios of our source (BHR71-IRS1; circle), IRAS 16293B (cross), IRAS 16293A (square), V883 Ori (upward triangle), G31.41 (downward triangle), and Sgr B2(N2b) (star). Each color represents a different ratio, and each symbol represents a different object.}
  \label{fig:comp_IRAS_Sgr}
\end{figure}

Isotopologues of methyl cyanide were detected around several low-mass objects (e.g., \citealt{Taquet2019}; \citealt{MartinDomenech2021}; \citealt{Nazari2021}; \citealt{Bianchi2022}; \citealt{Cruz2025}). However, we limited the comparison to only the objects that have also been observed with a similarly broad frequency range and at a similarly high sensitivity level with ALMA, i.e., Sgr B2(N2b) \citep{Belloche2025}, IRAS 16293A and B \citep{Calcutt2018}, G31.41 \citep{Mininni2023}, and V883 Ori \citep{Jeong2025}. Figure \ref{fig:comp_IRAS_Sgr} presents a comparison of methyl cyanide isotopologue ratios toward BHR71-IRS1, IRAS 16293A and B, V883 Ori, G31.41, and Sgr B2(N2b). We note that if a fixed source size is assumed when measuring column densities, the column density ratios of two molecules with different sublimation temperatures may not be the same as their abundance ratios (\citealt{Nazari2021}; \citealt{Nazari2024_scatter}). This is simply because the total hydrogen column density is different for the different regions that the two molecules trace. However, as the isotopologues of methyl cyanide are expected to have similar sublimation temperatures (supported by the similar $T_{\rm ex}$ found for them; see Table \ref{tab:results}), the column density ratios of these molecules in Fig. \ref{fig:comp_IRAS_Sgr} are likely similar to their abundance ratios. In other words, the variations in the column density ratios in Fig. \ref{fig:comp_IRAS_Sgr} are likely dominated by chemical factors. 

Figure \ref{fig:comp_IRAS_Sgr} shows that the $^{13}$CH$_3$CN/CH$_3^{13}$CN ratio toward BHR71-IRS1, IRAS 16293A and B, V883 Ori, G31.41, and Sgr B2(N2b) is consistent with unity, which points to the similar probability of $^{13}$C being incorporated in the CH$_3$ or CN group. Considering that all these objects point to a $^{13}$CH$_3$CN/CH$_3^{13}$CN ratio of unity, deviations from unity may be an indication of inaccuracy in a column density measurement.

Although the ratios of CH$_2$DCN/CH$_3$CN of the low-mass systems BHR71-IRS1, IRAS 16293A and B, and V883 Ori are all ${\sim}0.05-0.1$, it is around one order of magnitude lower for the high-mass Sgr B2(N2b). The same trend has been observed for CH$_2$DOH/CH$_3$OH in a large sample of low- and high-mass protostellar systems and was attributed to either the higher temperatures of high-mass star-forming regions or differences between timescales of low- and high-mass star formation processes (\citealt{Drozdovskaya2022}; \citealt{vangelder2022}; \citealt{Sakai2025}). The same might hold for methyl cyanide. It is worth mentioning that our CH$_2$DCN/CH$_3$CN value for BHR71-IRS1 is in good agreement with those found toward other objects in the literature, even those with narrower frequency bandwidth observations (\citealt{Taquet2015}; \citealt{Taquet2019}; \citealt{MartinDomenech2021}; \citealt{Bianchi2022}).

The ratio of CH$_3$C$^{15}$N/CH$_3$CN seems to be the highest toward BHR71-IRS1 at ${\sim}8 \times 10^{-3}$ (i.e., a CH$_3$CN/CH$_3$C$^{15}$N of ${\sim}130$). This ratio would be slightly lower, ${\sim}6 \times 10^{-3}$, for a fixed excitation temperature of 150\,K but still remains the highest among the sources presented in Fig. \ref{fig:comp_IRAS_Sgr}. On the other hand, this ratio is the lowest toward Sgr B2(N2b), at ${\sim}3 \times 10^{-3}$. Considering the error bars and the small source sample, no further conclusions can be made on the differences between objects. Future COMPASS works will be able to answer whether the variations are common across other nitrogen-bearing COMs and whether such a trend holds across various protostellar systems.

\section{D/H ratio}
\label{sec:like_methanol}

Using the doubly deuterated and singly deuterated isotopologues of CH$_3$CN and considering the statistical correction, D/H can be calculated as (see Appendix B of \citealt{Manigand2019} for more details)

\begin{align}
    \rm{D/H_{(single)}} &= \frac{1}{3} N({\rm{CH_2DCN}})/N({\rm{CH_3CN}}),\\
    \rm{D/H_{(double)}} &=  \sqrt{\frac{1}{3} N({\rm{CHD_2CN}})/N(\rm{CH_3CN})}\\
    \rm{D/H_{(double/single)}} &= N({\rm{CHD_2CN}})/N(\rm{CH_2DCN}).
    \label{eq:D_H}
\end{align}

\begin{figure}
  \resizebox{\columnwidth}{!}{\includegraphics{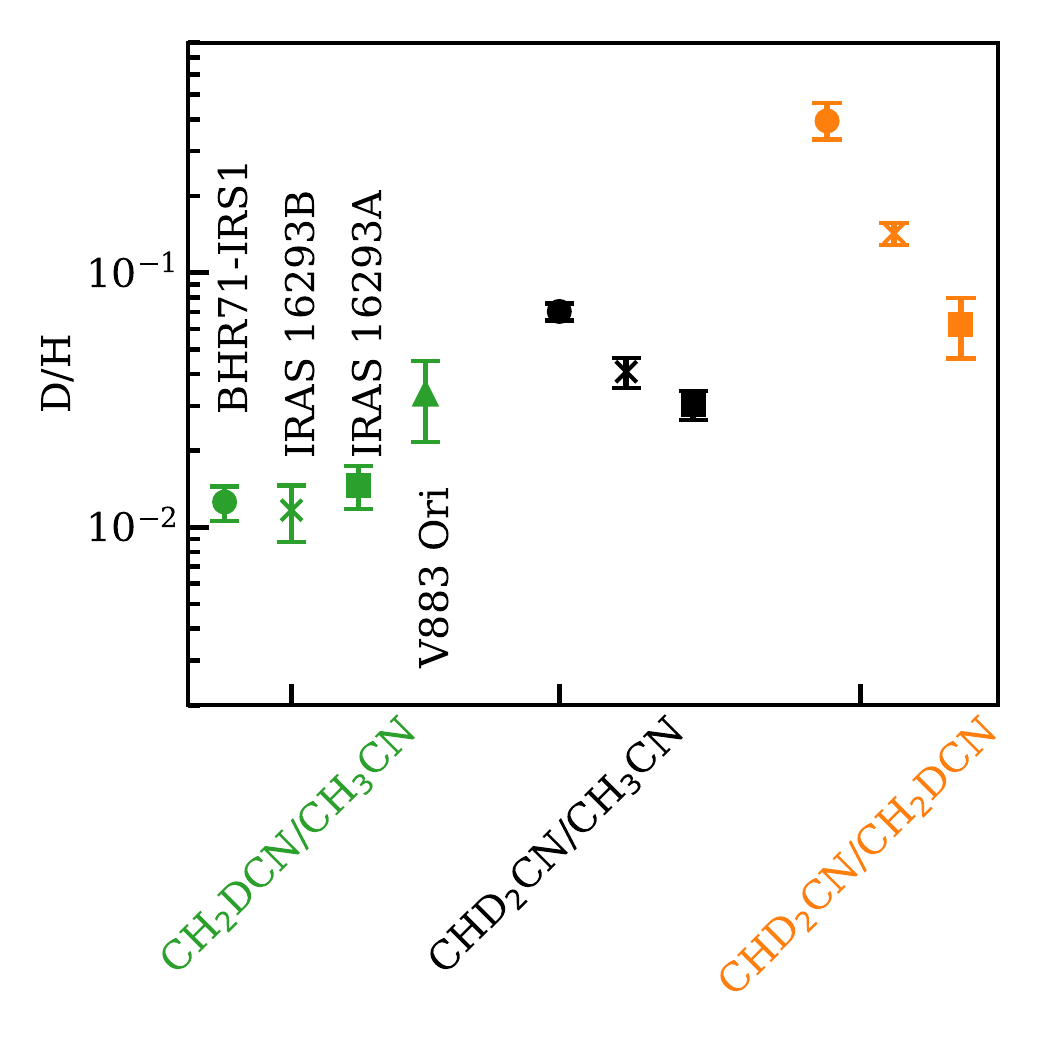}}
  \caption{D/H ratios calculated from different tracers after consideration of statistical corrections.}
  \label{fig:comp_D_H}
\end{figure}

\noindent The subscripts identify how the D/H is measured: through a singly deuterated isotopologue, a doubly deuterated isotopologue, or both doubly and singly deuterated ones. The D/H ratios found from methyl cyanide and methanol (\citealt{Coutens_COMPASS}) isotopologues are presented in Table \ref{tab:D_H}. The absolute value of D/H found for methanol toward BHR71-IRS1 from the analogous isotopologue ratios agrees within the uncertainties with the D/H found here for methyl cyanide. Figure \ref{fig:comp_D_H} presents these D/H ratios from different isotopologues of methyl cyanide for the low-mass protostellar systems considered in Fig. \ref{fig:comp_IRAS_Sgr}. The D/H from CH$_2$DCN/CH$_3$CN for various low-mass protostellar systems is in the range ${\sim}1-3$\%, placing CH$_3$CN in the group of objects with low deuteration in \cite{Jorgensen2018}, which included methanol. These similarities between methyl cyanide and methanol D/H values suggest that methyl cyanide can be formed and deuterated in a similar star formation stage as methanol. As most of methanol deuteration likely occurs in the cold dense core phase (\citealt{Jorgensen2018}), a similar conclusion can be made for CH$_3$CN. Comparing with chemical models, this would be in line with an early formation of methyl cyanide in the chemical models of \cite{Garrod2022} with a ``fast'' warm-up timescale, but not with the mostly late formation of this molecule in models with the ``slow'' or ``medium'' warm-up timescales.

\begin{table}
\centering
\renewcommand{\arraystretch}{1.2}
    \caption{D/H ratios from methyl cyanide and methanol toward the continuum peak of BHR71-IRS1.}
    \label{tab:D_H}
    \begin{tabular}{@{\extracolsep{1.2mm}}*{3}{l}}
          \toprule
          \toprule      
         & Methyl cyanide  &  Methanol   \\ 
        \midrule 
$\rm{D/H_{(single)}}$ & 0.013$\pm$0.002 & 0.019$\pm$0.006 \\
$\rm{D/H_{(double)}}$ & 0.070$_{-0.006}^{+0.005}$ & 0.100$\pm$0.020\\
$\rm{D/H_{(double/single)}}$ & 0.400$^{+0.070}_{-0.061}$ & 0.550$\pm$0.170\\

\bottomrule
        \end{tabular}
        \tablefoot{Methanol values are taken from \cite{Coutens_COMPASS}.}
\end{table}

Figure \ref{fig:comp_D_H} shows that the D/H ratio of BHR71-IRS1 varies by more than one order of magnitude depending on the tracer used for its measurement. In other words, D/H gets higher as the number of deuterium atoms increases in a molecule. This is similar to the trend that is observed for methanol toward the same object (see our Table \ref{tab:D_H} and \citealt{Coutens_COMPASS}), toward IRAS 16293B (\citealt{Drozdovskaya2022}), and in a large sample of high-mass protostellar systems (\citealt{vangelder2022}). A similar trend has also been observed for water (\citealt{Coutens2014}; \citealt{Jensen2021}) and COMs other than methanol, for example methyl formate \citep{Manigand2019}, dimethyl ether \citep{Richard2021}, acetaldehyde \citep{Ferrer2023}, and ethylene oxide \citep{Muller2023}. This trend may suggest that the more deuterated isotopologues form at later times during the pre-stellar phase when the temperatures are lower.

Furthermore, in Fig. \ref{fig:comp_D_H} the D/H ratios of the different low-mass objects based on CH$_2$DCN/CH$_3$CN and CHD$_2$CN/CH$_3$CN are similar (within a factor of ${\sim}3$). However, using CHD$_2$CN/CH$_2$DCN, there is a factor of ${\sim}6-10$ spread in D/H between the three low-mass protostellar systems. In particular, the D/H based on this latter ratio is a factor of ${\sim}3$ higher toward BHR71-IRS1 than IRAS 16293B and a factor of ${\sim}10$ higher than IRAS 16293A, suggesting that the pre-stellar temperatures may have got lower for BHR71-IRS1 than for IRAS 16293A and B. Interestingly, BHR71-IRS1 also shows a larger difference in D/H compared with IRAS 16293A and B when measured via CHD$_2$OH/CH$_2$DOH as opposed to CH$_2$DOH/CH$_3$OH or CHD$_2$OH/CH$_3$OH (\citealt{Coutens_COMPASS}). Considering other literature measurements, the value of CHD$_2$CN/CH$_2$DCN toward BHR71-IRS1 is similar to that found toward B1-c (\citealt{Nazari2021}). This is in line with the spread seen in CHD$_2$CN/CH$_2$DCN across the three low-mass systems in Fig. \ref{fig:comp_D_H}.

\section{Conclusions}
\label{sec:conclusions}

This work used COMPASS data of BHR71-IRS1 to quantify the column densities of methyl cyanide isotopologues in a wide frequency range (${\sim}33$\,GHz). Our main conclusions are summarized below.

\begin{itemize}
    \item From the isotopologues with available spectroscopy, we detected CH$_3$CN, $^{13}$CH$_3$CN, CH$_3^{13}$CN, CH$_3$C$^{15}$N, CH$_2$DCN, and CHD$_2$CN.\ We did not detect $^{13}$CH$_3^{13}$CN or CD$_3$CN, which is consistent with their expected low abundances. 
    \item The ratio of $^{13}$CH$_3$CN/CH$_3^{13}$CN is consistent with unity and similar for BHR71-IRS1, IRAS 16293A and B, V883 Ori, G31.41, and Sgr B2(N2b).
    \item The CH$_2$DCN/CH$_3$CN ratio is consistent between BHR71-IRS1, IRAS 16293A and B, and V883 Ori but is around one order of magnitude higher than that of Sgr B2(N2b), which might be related to the warmer pre-stellar phase or different formation timescale of Sgr B2(N2b). 
    \item The D/H ratios found from different methyl cyanide isotopologues agree well with the analogous measurements carried out using methanol toward BHR71-IRS1. This may point to the formation and deuteration of methyl cyanide at a similar stage as methanol during star formation (i.e., the pre-stellar phase).
    \item The D/H ratio for methyl cyanide increases with the number of deuterium atoms. The same trend is observed for methanol in BHR71-IRS1 and could point to the deuterated isotopologues forming at a later and colder point in the pre-stellar stage. 
\end{itemize}

A larger sample of protostellar systems with similar frequency coverage and sensitivity is needed to confirm the trends observed here statistically. This will be further investigated by future COMPASS studies.

\begin{acknowledgements}
     We are deeply grateful to everyone at the ALMA observatory that enabled the successful execution of this Large Program and, in particular, staff at the Nordic ALMA Regional Centre node for their support during its planning and execution stages. This paper makes use of the following ALMA data: ADS/JAO.ALMA\#2022.1.00316.L. ALMA is a partnership of ESO (representing its member states), NSF (USA) and NINS (Japan), together with NRC (Canada), MOST and ASIAA (Taiwan), and KASI (Republic of Korea), in cooperation with the Republic of Chile. The Joint ALMA Observatory is operated by ESO, AUI/NRAO and NAOJ. The National Radio Astronomy Observatory is a facility of the National Science Foundation operated under cooperative agreement by Associated Universities, Inc. The project leading to this publication has received support from ORP, that is funded by the European Union’s Horizon 2020 research and innovation programme under grant agreement No 101004719 [ORP]. P.N. acknowledges support from the ESO Fellowship and the IAU Gruber Foundation Fellowship programs. A.C. and F.C.S.M. received financial support from the European Research Council (ERC) under the European Union’s Horizon 2020 research and innovation programme (ERC Starting Grant “Chemtrip”, grant agreement No 949278). S.M. acknowledges support from the Thematic Action``Physique et Chimie du Milieu Interstellaire'' (PCMI) of INSU Programme National ``Astro'', with contributions from CNRS Physique \& CNRS Chimie, CEA, and CNES. J.-E.L., J.-H.J., H.-S.Y., and C.-H.K. were supported by the National Research Foundation of Korea (NRF) grant funded by the Korea government (MSIT) (grant numbers RS-2026-25490557 and RS-2024-00416859). J.-H.J was also supported by Basic Science Research Program through the National Research Foundation of Korea (NRF) funded by the Ministry of Education (grant number RS-2025-25402622). D.H. is supported by a Center for Informatics and Computation in Astronomy (CICA) grant and grant number 110J0353I9 from the Ministry of Education of Taiwan. D.H. also acknowledges support from the National Science and Technology Council, Taiwan (Grant NSTC111-2112-M-007-014-MY3, NSTC113-2639-M-A49-002-ASP, and NSTC113-2112-M-007-027). J. Ferrer Asensio and S. Zeng thank RIKEN Special Postdoctoral Researcher Program (Fellowships) for financial support. A.K. acknowledges support by the NKFIH NKKP grant ADVANCED 149943 and the NKFIH excellence grant TKP2021-NKTA-64. Project no.149943 has been implemented with the support provided by the Ministry of Culture and Innovation of Hungary from the National Research, Development and Innovation Fund, financed under the NKKP ADVANCED funding scheme. A.K. also acknowledges the work from COST Action PLANETS CA22133, supported by COST (European Cooperation in Science and Technology). S.S. acknowledges the support of the Max Planck Society. Y.-L. Yang acknowledges support from a Grant-in-Aid from the Ministry of Education, Culture, Sports, Science, and Technology of Japan (20H05845, 20H05844), and a pioneering project in RIKEN (Evolution of Matter in the Universe). 
\end{acknowledgements}


\bibliographystyle{aa}
\bibliography{aa58748-25}

\begin{appendix}

\section{Transitions covered in the data}
\label{app:transitions}
The data are taken from the CDMS database. Most of the data for CH$_3$CN and CH$_3$CN $v_8$=1 were reported in \cite{Muller2009,Muller2015} and \citet[also see \citealt{Kukolich1973}; \citealt{Boucher1977}; \citealt{Kukolich1982}; \citealt{Anttila1993}]{Cazzoli2006}. The spectroscopic data for $^{13}$CH$_3$CN and CH$_3^{13}$CN in the CDMS were taken from \cite{Pearson1996}, \cite{Muller2009}, \cite{Demaison1979}, and \cite{Kukolich1982}. The data for CH$_3$C$^{15}$N are given in \cite{Muller2016} and those for $^{13}$CH$_3^{13}$CN were taken from \cite{Muller2009}. The data for CH$_2$DCN and CHD$_2$CN in the CDMS are given in \cite{Nguyen2013} and \cite{Muller2009}. Those of CD$_3$CN are taken from \cite{Demaison1979} and \cite{Brown1989}. Tables \ref{tab:transition_CH3CN}-\ref{tab:transition_CHD2CN} present all the transitions that are covered and detected in the data. The limits put on $E_{\rm up}$ and $A_{\rm ij}$ are the same as those in Figs. \ref{fig:13CH3CN} and \ref{fig:CH3C-13-N}-\ref{fig:CHD2CN}.

\begin{table}[h]
    \caption{Transitions of CH$_3$CN.}
    \label{tab:transition_CH3CN}
    \setlength{\tabcolsep}{5pt}
    \resizebox{\columnwidth}{!}{\begin{tabular}{*{5}{l}}
          \toprule
          \toprule      
     Transition & Frequency & $A_{\rm ij}$ & $E_\mathrm{up}$ & Possible  \\
   & (MHz) & (s$^{-1}$) & (K) &   blends\\
        \midrule 
 16 11 - 15 11 & 293\,620.295 & $1.16 \times 10^{-3}$ & 982.5 & --\\
 $^*$16 10 - 15 10 & 293\,738.276 & $1.34 \times 10^{-3}$ & 833.1 & --\\
 $^*$16 9 - 15 -9 & 293\,845.164 & $1.51 \times 10^{-3}$ & 697.8 & --\\
 $^*$16 -9 - 15 9 & 293\,845.164 & $1.51 \times 10^{-3}$ & 697.8 & --\\
 16 8 - 15 8 & 293\,940.917 & $1.65 \times 10^{-3}$ & 576.7 & --\\
 16 7 - 15 7 & 294\,025.496 & $1.78 \times 10^{-3}$ & 469.8 & CH$_3$CHO\\
 16 6 - 15 -6 & 294\,098.867 & $1.90 \times 10^{-3}$ & 377.1 & --\\
 16 -6 - 15 6 & 294\,098.867 & $1.90 \times 10^{-3}$ & 377.1 & --\\
 16 5 - 15 5 & 294\,161.002 & $1.99 \times 10^{-3}$ & 298.6 & -- \\
 16 4 - 15 4 & 294\,211.873 & $2.07 \times 10^{-3}$ & 234.3 & --\\
 16 3 - 15 -3 & 294\,251.461 & $2.13 \times 10^{-3}$ & 184.4 & --\\
 16 -3 - 15 3 & 294\,251.461 & $2.13 \times 10^{-3}$ & 184.4 & --\\
 16 2 - 15 2 & 294\,279.75 & $2.18 \times 10^{-3}$ & 148.6 & --\\
 16 1 - 15 1 & 294\,296.728 & $2.20 \times 10^{-3}$ & 127.2 & --\\
 16 0 - 15 0 & 294\,302.389 & $2.21 \times 10^{-3}$ & 120.1 & --\\
\bottomrule
        \end{tabular}}
        \tablefoot{Only transitions with $E_{\rm up}< 1000$\,K and $A_{\rm ij} > 5\times 10^{-7}$\,s$^{-1}$ are shown. The transitions that are detected and used for the fitting are marked with a star. The rest are blended, extremely optically thick, or not detected.}
\end{table}

\begin{table}[h]
    \caption{Transitions of $^{13}$CH$_3$CN.}
    \label{tab:transition_13CH3CN}
    \setlength{\tabcolsep}{3pt}
    \resizebox{\columnwidth}{!}{\begin{tabular}{*{5}{l}}
          \toprule
          \toprule      
    Transition & Frequency & $A_{\rm ij}$ & $E_\mathrm{up}$ & Possible \\
 & (MHz) & (s$^{-1}$) & (K) & blends \\
        \midrule 
 16 9 - 15 9 & 285\,372.864 & $1.38 \times 10^{-3}$ & 695.4 & -- \\
 16 8 - 15 8 & 285\,463.677 & $1.51 \times 10^{-3}$ & 574.1 & --\\
 16 7 - 15 7 & 285\,543.893 & $1.63 \times 10^{-3}$ & 467.0 & --\\
 16 6 - 15 6 & 285\,613.481 & $1.74 \times 10^{-3}$ & 374.1 & CH$_3$OCHO, C$_2$H$_5$OH\\
 $^*$16 5 - 15 5 & 285\,672.411 & $1.83 \times 10^{-3}$ & 295.4& --\\
 16 4 - 15 4 & 285\,720.66 & $1.90 \times 10^{-3}$ & 231.1 & HNCO, CH$_3$COCH$_3$, \\&&&&C$_2$H$_5$OH, CH$_3$OCHO, \\&&&&aGg'-(CH$_2$OH)$_2$ \\
 $^*$16 3 - 15 3 & 285\,758.207 & $1.95 \times 10^{-3}$ & 181.0& --\\
 $^*$16 2 - 15 2 & 285\,785.038 & $1.99 \times 10^{-3}$ & 145.2& --\\
 $^*$16 1 - 15 1 & 285\,801.14 & $2.02 \times 10^{-3}$ & 123.8& --\\
 $^*$16 0 - 15 0 & 285\,806.509 & $2.03 \times 10^{-3}$ & 116.6& --\\
 17 8 - 16 8 & 303\,297.154 & $1.89 \times 10^{-3}$ & 588.6& --\\
 17 7 - 16 7 & 303\,382.353 & $2.02 \times 10^{-3}$ & 481.5& --\\
 17 6 - 16 6 & 303\,456.263 & $2.13 \times 10^{-3}$ & 388.6 & NH$_2$CHO, CH$_3$SH\\
 17 5 - 16 5 & 303\,518.854 & $2.22 \times 10^{-3}$ & 310.0 & Unidentified\\
 17 4 - 16 4 & 303\,570.099 & $2.30 \times 10^{-3}$ & 245.6 & CHD$_2$OH, CH$_3$SH, \\&&&&CD$_3$OH, s-propanal\\
 $^*$17 3 - 16 3 & 303\,609.971 & $2.36 \times 10^{-3}$ & 195.6 & --\\
 $^*$17 2 - 16 2 & 303\,638.482 & $2.40 \times 10^{-3}$ & 159.8 & --\\
 $^*$17 1 - 16 1 & 303\,655.577 & $2.43 \times 10^{-3}$ & 138.3 & --\\
 $^*$17 0 - 16 0 & 303\,661.278 & $2.43 \times 10^{-3}$ & 131.2 & --\\
\bottomrule
        \end{tabular}}
        \tablefoot{Only transitions with $E_{\rm up}< 700$\,K and $A_{\rm ij} > 10^{-6}$\,s$^{-1}$ are shown. The transitions that are detected and used for the fitting are marked with a star. The rest are either blended or not detected.}
\end{table}

\begin{table}[h]
    \caption{Transitions of CH$_3^{13}$CN.}
    \label{tab:transition_CH3C-13-N}
    \setlength{\tabcolsep}{5pt}
    \resizebox{\columnwidth}{!}{\begin{tabular}{*{5}{l}}
          \toprule
          \toprule      
     Transition & Frequency & $A_{\rm ij}$ & $E_\mathrm{up}$ & Possible\\
   & (MHz) & (s$^{-1}$) & (K) &  blends \\
        \midrule 
 16 9 - 15 9 & 293\,701.434 & $1.50 \times 10^{-3}$ & 697.8 & --\\
 16 8 - 15 8 & 293\,796.795 & $1.65 \times 10^{-3}$ & 576.7 & --\\
 16 7 - 15 7 & 293\,881.027 & $1.78 \times 10^{-3}$ & 469.7 & Unidentified\\
 16 6 - 15 6 & 293\,954.097 & $1.89 \times 10^{-3}$ & 377.0 & CH$_3$OCHO\\
 $^*$16 5 - 15 5 & 294\,015.975 & $1.99 \times 10^{-3}$ & 298.5 & --\\
 $^*$16 4 - 15 4 & 294\,066.637 & $2.07 \times 10^{-3}$ & 234.3 & --\\
 $^*$16 3 - 15 3 & 294\,106.059 & $2.13 \times 10^{-3}$ & 184.3 & --\\
 $^*$16 2 - 15 2 & 294\,134.23 & $2.17 \times 10^{-3}$ & 148.6 & --\\
 $^*$16 1 - 15 1 & 294\,151.143 & $2.20 \times 10^{-3}$ & 127.2 & --\\
 16 0 - 15 0 & 294\,156.78 & $2.21 \times 10^{-3}$ & 120.0 & --\\
\bottomrule
        \end{tabular}}
        \tablefoot{Only transitions with $E_{\rm up}< 700$\,K and $A_{\rm ij} > 10^{-6}$\,s$^{-1}$ are shown. The transitions that are detected and used for the fitting are marked with a star. The rest are either blended or not detected.}
\end{table}

\begin{table}[h]
    \caption{Transitions of CH$_3$CN $v_8$=1.}
    \label{tab:transition_CH3CN_v8}
    \setlength{\tabcolsep}{3pt}
    \resizebox{\columnwidth}{!}{\begin{tabular}{*{5}{l}}
          \toprule
          \toprule      
     Transition & Frequency & $A_{\rm ij}$ & $E_\mathrm{up}$ & Possible \\
 & (MHz) & (s$^{-1}$) & (K) & blends \\
        \midrule 
 16 6 2 - 15 6 2 & 294\,899.269 & $1.90 \times 10^{-3}$ & 982.1 & --\\
 $^*$16 -1 3 - 15 1 3 & 294\,904.428 & $2.20 \times 10^{-3}$ & 639.3 & --\\
 16 8 3 - 15 8 3 & 294\,921.109 & $1.65 \times 10^{-3}$ & 995.7 & --\\
 16 5 2 - 15 -5 2 & 294\,973.178 & $1.99 \times 10^{-3}$ & 890.4 & CH$_3$OCHO\\
 16 -5 2 - 15 5 2 & 294\,973.178 & $1.99 \times 10^{-3}$ & 890.4 & CH$_3$OCHO\\
 16 7 3 - 15 -7 3 & 294\,994.776 & $1.79 \times 10^{-3}$ & 902.1 & CH$_3$OCHO\\
 16 -7 3 - 15 7 3 & 294\,994.776 & $1.79 \times 10^{-3}$ & 902.1 & CH$_3$OCHO\\
 16 4 2 - 15 4 2 & 295\,035.501 & $2.07 \times 10^{-3}$ & 813.0 & CHD$_2$OH\\
 16 6 3 - 15 6 3 & 295\,057.433 & $1.90 \times 10^{-3}$ & 822.7 & CH$_2$DCN\\
 16 3 2 - 15 3 2 & 295\,085.98 & $2.13 \times 10^{-3}$ & 749.7 & Unidentified\\
 $^*$16 5 3 - 15 5 3 & 295\,109.283 & $1.99 \times 10^{-3}$ & 757.5 & --\\
 $^*$16 2 2 - 15 -2 2 & 295\,124.061 & $2.18 \times 10^{-3}$ & 700.8 & --\\
 $^*$16 -2 2 - 15 2 2 & 295\,124.061 & $2.18 \times 10^{-3}$ & 700.8 & --\\
 $^*$16 1 2 - 15 1 2 & 295\,148.141 & $2.20 \times 10^{-3}$ & 666.0 & --\\
 $^*$16 0 2 - 15 0 2 & 295\,150.385 & $2.21 \times 10^{-3}$ & 645.6 & --\\
 $^*$16 4 3 - 15 -4 3 & 295\,150.836 & $2.07 \times 10^{-3}$ & 706.6 & --\\
 $^*$16 -4 3 - 15 4 3 & 295\,150.836 & $2.07 \times 10^{-3}$ & 706.6 & --\\
 $^*$16 3 3 - 15 3 3 & 295\,183.66 & $2.13 \times 10^{-3}$ & 669.9 & --\\
 $^*$16 2 3 - 15 2 3 & 295\,215.569 & $2.18 \times 10^{-3}$ & 647.6 & --\\
 $^*$16 1 3 - 15 -1 3 & 295\,472.912 & $2.21 \times 10^{-3}$ & 639.6 & --\\
\bottomrule
        \end{tabular}}
        \tablefoot{Only transitions with $E_{\rm up}< 1000$\,K and $A_{\rm ij} > 10^{-6}$\,s$^{-1}$ are shown. The transitions that are detected and used for the fitting are marked with a star. The rest are either blended or not detected.}
\end{table}

\begin{table}[h]
    \caption{Transitions of CH$_3$C$^{15}$N.}
    \label{tab:transition_CH3CN-15}
    \setlength{\tabcolsep}{3pt}
    \resizebox{\columnwidth}{!}{\begin{tabular}{*{5}{l}}
          \toprule
          \toprule      
     Transition & Frequency & $A_{\rm ij}$ & $E_\mathrm{up}$ & Possible \\
   & (MHz) & (s$^{-1}$) & (K) & blends \\
        \midrule 
 $^*$16 6 - 15 6 & 285\,253.161 & $1.73 \times 10^{-3}$ & 373.9 & --\\
 16 5 - 15 5 & 285\,312.336 & $1.82 \times 10^{-3}$ & 295.3 & CHD$_2$OH\\
 16 4 - 15 4 & 285\,360.784 & $1.89 \times 10^{-3}$ & 230.9 & CH$_3$OCHO\\
 $^*$16 3 - 15 3 & 285\,398.486 & $1.95 \times 10^{-3}$ & 180.9 & --\\
 $^*$16 2 - 15 2 & 285\,425.427 & $1.99 \times 10^{-3}$ & 145.1 & --\\
 $^*$16 1 - 15 1 & 285\,441.595 & $2.01 \times 10^{-3}$ & 123.6 & --\\
 $^*$16 0 - 15 0 & 285\,446.986 & $2.02 \times 10^{-3}$ & 116.5 & --\\
 $^*$17 6 - 16 6 & 303\,073.581 & $2.12 \times 10^{-3}$ & 388.5 & --\\
 17 5 - 16 5 & 303\,136.432 & $2.22 \times 10^{-3}$ & 309.8 & CH$_2$DOH, CH$_3$COCH$_3$\\
 17 4 - 16 4 & 303\,187.889 & $2.29 \times 10^{-3}$ & 245.5 & CH$_3$SH\\
 $^*$17 3 - 16 3 & 303\,227.936 & $2.35 \times 10^{-3}$ & 195.4 & --\\
 $^*$17 2 - 16 2 & 303\,256.554 & $2.40 \times 10^{-3}$ & 159.6 & --\\
 17 1 - 16 1 & 303\,273.73 & $2.42 \times 10^{-3}$ & 138.2 & CH$_3$SH\\
 17 0 - 16 0 & 303\,279.456 & $2.43 \times 10^{-3}$ & 131.0 & CH$_3$SH\\
\bottomrule
        \end{tabular}}
        \tablefoot{Only transitions with $E_{\rm up}< 400$\,K and $A_{\rm ij} > 10^{-3}$\,s$^{-1}$ are shown. The transitions that are detected and used for the fitting are marked with a star. The rest are either blended or not detected.}
\end{table}

\begin{table}[h]
    \caption{Transitions of CH$_2$DCN.}
    \label{tab:transition_CH2DCN}
    \setlength{\tabcolsep}{3pt}
    \resizebox{\columnwidth}{!}{\begin{tabular}{*{5}{l}}
          \toprule
          \toprule      
    Transition & Frequency & $A_{\rm ij}$ & $E_\mathrm{up}$ & Possible \\
 & (MHz) & (s$^{-1}$) & (K) & blends \\
        \midrule 
 $^*$17 1 17 - 16 1 16 & 293\,869.52 & $2.20 \times 10^{-3}$ & 132.3 & --\\
 17 10 7 - 16 10 6 & 294\,699.57 & $1.45 \times 10^{-3}$ & 665.6 & --\\
 17 10 8 - 16 10 7 & 294\,699.57 & $1.45 \times 10^{-3}$ & 665.6 & --\\
 17 9 8 - 16 9 7 & 294\,792.16 & $1.60 \times 10^{-3}$ & 563.6 & CH$_3$OCHO\\
 17 9 9 - 16 9 8 & 294\,792.16 & $1.60 \times 10^{-3}$ & 563.6 & CH$_3$OCHO\\
 $^*$17 8 9 - 16 8 8 & 294\,875.03 & $1.73 \times 10^{-3}$ & 472.2 & --\\
 $^*$17 8 10 - 16 8 9 & 294\,875.03 & $1.73 \times 10^{-3}$ & 472.2 & --\\
 17 7 10 - 16 7 9 & 294\,948.46 & $1.85 \times 10^{-3}$ & 391.5 & CH$_3$OH\\
 17 7 11 - 16 7 10 & 294\,948.46 & $1.85 \times 10^{-3}$ & 391.5 & CH$_3$OH\\
 $^*$17 6 11 - 16 6 10 & 295\,012.55 & $1.95 \times 10^{-3}$ & 321.5 & --\\
 $^*$17 6 12 - 16 6 11 & 295\,012.55 & $1.95 \times 10^{-3}$ & 321.5 & --\\
 $^*$17 0 17 - 16 0 16 & 295\,060.16 & $2.23 \times 10^{-3}$ & 127.5 & --\\
 $^*$17 5 12 - 16 5 11 & 295\,067.51 & $2.04 \times 10^{-3}$ & 262.2 & --\\
 $^*$17 5 13 - 16 5 12 & 295\,067.51 & $2.04 \times 10^{-3}$ & 262.2 & --\\
 $^*$17 4 14 - 16 4 13 & 295\,114.36 & $2.11 \times 10^{-3}$ & 213.8 & --\\
 $^*$17 4 13 - 16 4 12 & 295\,114.36 & $2.11 \times 10^{-3}$ & 213.8 & --\\
 $^*$17 2 16 - 16 2 15 & 295\,144.02 & $2.20 \times 10^{-3}$ & 149.1 & --\\
 $^*$17 3 15 - 16 3 14 & 295\,155.02 & $2.16 \times 10^{-3}$ & 176.0 & --\\
 $^*$17 3 14 - 16 3 13 & 295\,156.1 & $2.16 \times 10^{-3}$ & 176.0 & --\\
 $^*$17 2 15 - 16 2 14 & 295\,267.06 & $2.21 \times 10^{-3}$ & 149.1 & --\\
 $^*$17 1 16 - 16 1 15 & 296\,426.45 & $2.26 \times 10^{-3}$ & 133.5 & --\\
 $^*$18 1 18 - 17 1 17 & 311\,143.61 & $2.61 \times 10^{-3}$ & 147.3 & --\\
\bottomrule
        \end{tabular}}
        \tablefoot{Only transitions with $E_{\rm up}< 700$\,K and $A_{\rm ij} > 10^{-4}$\,s$^{-1}$ are shown. The transitions that are detected and used for the fitting are marked with a star. The rest are either blended or not detected.}
\end{table}

\begin{table}[h]
    \caption{Transitions of CHD$_2$CN.}
    \label{tab:transition_CHD2CN}
    \setlength{\tabcolsep}{3pt}
    \resizebox{\columnwidth}{!}{\begin{tabular}{*{5}{l}}
          \toprule
          \toprule      
    Transition & Frequency & $A_{\rm ij}$ & $E_\mathrm{up}$ & Possible \\
 & (MHz) & (s$^{-1}$) & (K) & blends \\
        \midrule 
 17 8 9 - 16 8 8 & 279\,903.981 & $1.48 \times 10^{-3}$ & 391.8 & CD$_3$OH, \\&&&& aGg'-(CH$_2$OH)$_2$\\
 17 8 10 - 16 8 9 & 279\,903.981 & $1.48 \times 10^{-3}$ & 391.8 & CD$_3$OH, \\&&&& aGg'-(CH$_2$OH)$_2$\\
 17 7 10 - 16 7 9 & 279\,969.203 & $1.58 \times 10^{-3}$ & 328.4 & CHD$_2$OH, CH$_3$OCHO\\
 17 7 11 - 16 7 10 & 279\,969.203 & $1.58 \times 10^{-3}$ & 328.4 & CHD$_2$OH, CH$_3$OCHO\\
 $^*$17 0 17 - 16 0 16 & 280\,008.184 & $1.91 \times 10^{-3}$ & 121.0 & --\\
 17 6 11 - 16 6 10 & 280\,026.312 & $1.67 \times 10^{-3}$ & 273.4 & Unidentified\\
 17 6 12 - 16 6 11 & 280\,026.312 & $1.67 \times 10^{-3}$ & 273.4 & Unidentified\\
 $^*$17 5 12 - 16 5 11 & 280\,075.75 & $1.74 \times 10^{-3}$ & 226.8 & --\\
 $^*$17 5 13 - 16 5 12 & 280\,075.75 & $1.74 \times 10^{-3}$ & 226.8 & --\\
 $^*$17 4 14 - 16 4 13 & 280\,118.627 & $1.80 \times 10^{-3}$ & 188.8 & --\\
 $^*$17 4 13 - 16 4 12 & 280\,118.627 & $1.80 \times 10^{-3}$ & 188.8 & --\\
 17 2 16 - 16 2 15 & 280\,131.335 & $1.88 \times 10^{-3}$ & 138.0 & C$_2$H$_5$CN\\
 17 3 15 - 16 3 14 & 280\,157.517 & $1.85 \times 10^{-3}$ & 159.1 & Unidentified \\
 17 3 14 - 16 3 13 & 280\,159.504 & $1.85 \times 10^{-3}$ & 159.1 & Unidentified \\
 $^*$17 2 15 - 16 2 14 & 280\,298.235 & $1.89 \times 10^{-3}$ & 138.0 & --\\
 $^*$17 1 16 - 16 1 15 & 281\,450.723 & $1.93 \times 10^{-3}$ & 125.8 & --\\
 $^*$18 1 18 - 17 1 17 & 295\,196.288 & $2.23 \times 10^{-3}$ & 138.8 & --\\
 $^*$18 7 11 - 17 7 10 & 296\,430.692 & $1.92 \times 10^{-3}$ & 342.6 & --\\
 $^*$18 7 12 - 17 7 11 & 296\,430.692 & $1.92 \times 10^{-3}$ & 342.6 & --\\
 $^*$18 0 18 - 17 0 17 & 296\,450.115 & $2.27 \times 10^{-3}$ & 135.2 & --\\
 18 6 12 - 17 6 11 & 296\,491.303 & $2.02 \times 10^{-3}$ & 287.6 & Unidentified\\
 18 6 13 - 17 6 12 & 296\,491.303 & $2.02 \times 10^{-3}$ & 287.6 & Unidentified\\
 18 5 14 - 17 5 13 & 296\,543.923 & $2.09 \times 10^{-3}$ & 241.1 & SO, CH$_2$DCHO,\\&&&& SO$_2$, CH$_3$OCHO,\\&&&& C$_2$H$_5$CN\\
 18 5 13 - 17 5 12 & 296\,543.923 & $2.09 \times 10^{-3}$ & 241.1 & SO, CH$_2$DCHO,\\&&&& SO$_2$, CH$_3$OCHO,\\&&&& C$_2$H$_5$CN\\
 $^*$18 4 15 - 17 4 14 & 296\,589.82 & $2.16 \times 10^{-3}$ & 203.0 & --\\
 $^*$18 4 14 - 17 4 13 & 296\,589.82 & $2.16 \times 10^{-3}$ & 203.0 & --\\
 $^*$18 2 17 - 17 2 16 & 296\,598.26 & $2.24 \times 10^{-3}$ & 152.2 & --\\
 $^*$18 3 16 - 17 3 15 & 296\,632.013 & $2.21 \times 10^{-3}$ & 173.4 & --\\
 $^*$18 3 15 - 17 3 14 & 296\,634.639 & $2.21 \times 10^{-3}$ & 173.4 & --\\
 $^*$18 2 16 - 17 2 15 & 296\,796.27 & $2.25 \times 10^{-3}$ & 152.2 & --\\
 $^*$18 1 17 - 17 1 16 & 297\,993.055 & $2.30 \times 10^{-3}$ & 140.1 & --\\
 19 1 19 - 18 1 18 & 311\,581.794 & $2.63 \times 10^{-3}$ & 153.8 & CD$_3$OH, C$_2$H$_5$OH,\\&&&& s-propanal\\
\bottomrule
        \end{tabular}}
        \tablefoot{Only transitions with $E_{\rm up}< 400$\,K and $A_{\rm ij} > 10^{-3}$\,s$^{-1}$ are shown. The transitions that are detected and used for the fitting are marked with a star. The rest are either blended or not detected.}
\end{table}

\clearpage

\section{Fitting methanol and methyl cyanide lines in the image plane}
\label{app:fit_size}

\begin{figure*}[b]
    \centering
    \includegraphics[width=\textwidth]{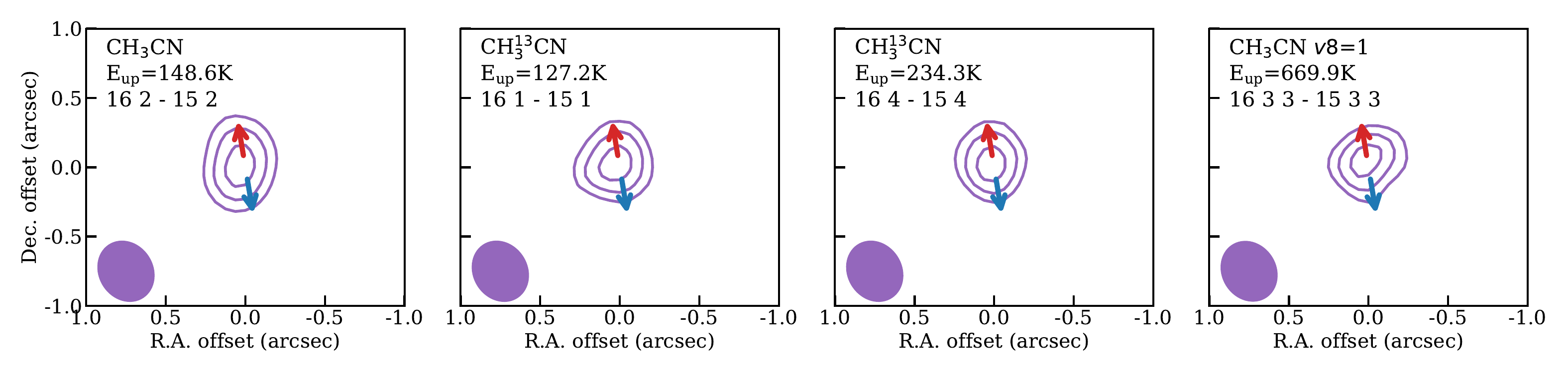}
    \caption{Integrated intensity (moment 0) maps of a few lines of minor and major isotopologues of methyl cyanide shown as contours with levels at 50\%, 70\%, and 90\% of the peak of each map. The lines from left to right have frequencies of 294.280, 294.151, 294.067, and 295.184\,GHz. The lines are integrated over 16 , 9, 7, and 10\,km\,s$^{-1}$ to ensure full coverage while minimizing contamination from blended species. The beam is highlighted on the bottom left. The blue and redshifted outflow are indicated by the blue and red arrows, respectively. The emission is not fully resolved. Thus, little difference is found between emitting areas of lines with vastly different upper-level energies.} 
    \label{fig:maps}
\end{figure*}

\begin{figure}
    \centering
    \includegraphics[width=0.7\columnwidth]{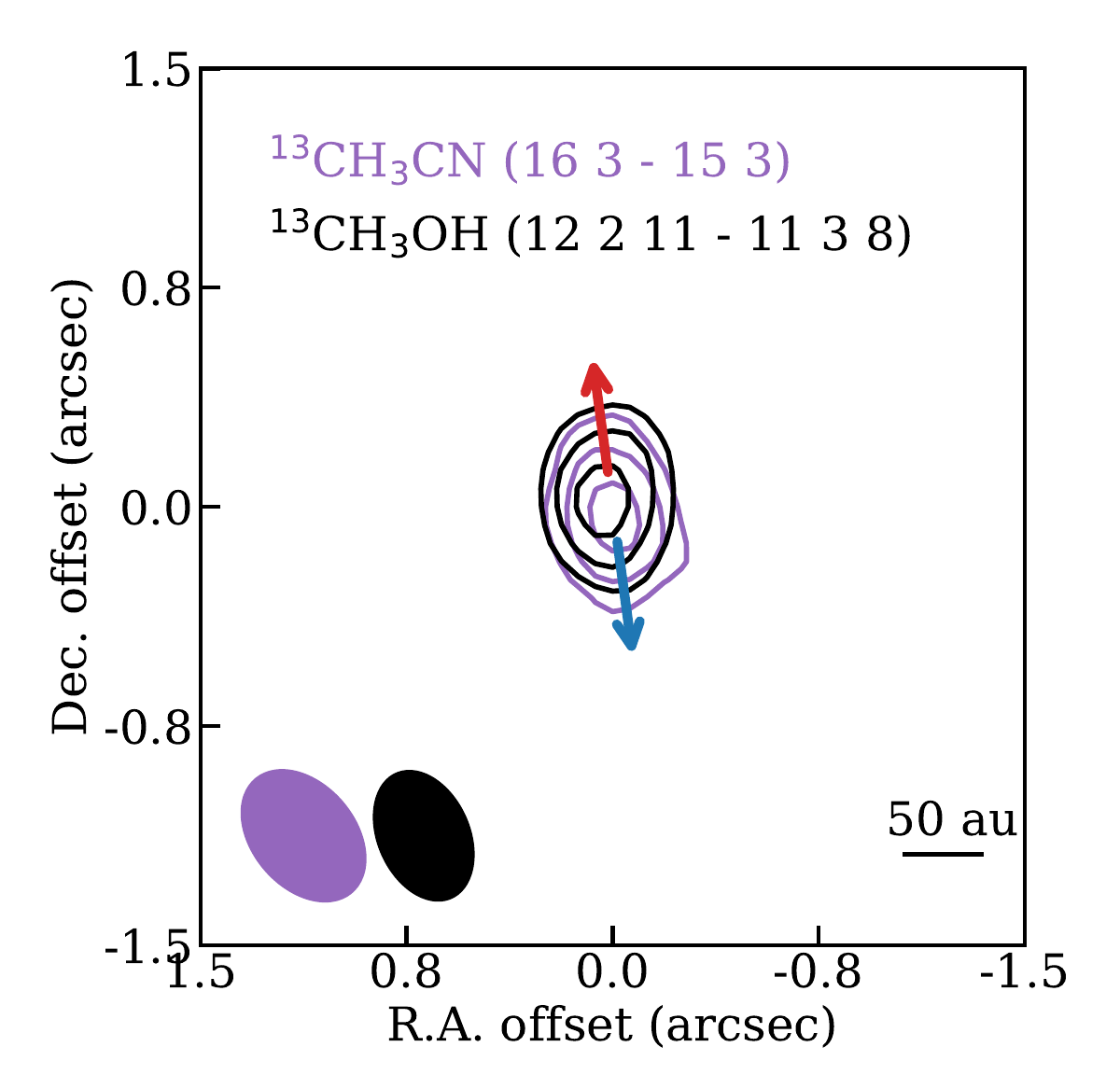}
    \caption{Comparison of emitting regions of $^{13}$CH$_3$CN and $^{13}$CH$_3$OH. Moment 0 maps of $^{13}$CH$_3$CN (purple; $\nu$=285.758\,GHz, $E_{\rm up}$=181.0\,K) and $^{13}$CH$_3$OH (black; $\nu$=310.147\,GHz, $E_{\rm up}$=214.4\,K) are shown as contours with levels at 50\%, 70\%, and 90\% of the peak of each map (roughly at around 0.1\,Jy\,beam$^{-1}$\,km\,s$^{-1}$). The two lines of $^{13}$CH$_3$CN and $^{13}$CH$_3$OH are integrated over 10 and 6\,km\,s$^{-1}$ to include the entire line without including any blended lines. The arrows show the red and blueshifted outflow direction. The two molecules show similar emitting regions, even though bulk of the emission (i.e., contours at $\geq 50\%$ of the peak) is not spatially resolved.}
    \label{fig:size_fitting}
\end{figure}

Figure \ref{fig:maps} presents the moment 0 maps of a few methyl cyanide lines with varying upper-level energies. Not much difference is seen between the lines regardless of their upper-level energy. This is likely due to the emission not being fully resolved. Figure \ref{fig:size_fitting} presents moment 0 maps of two lines with similar upper energy levels from $^{13}$CH$_3$CN and $^{13}$CH$_3$OH. This figure shows that the two lines trace similar regions within the limitations of our beam.

\clearpage
\section{Fitted spectra}

\begin{figure*}[b]
    \centering
    \includegraphics[width=\textwidth]{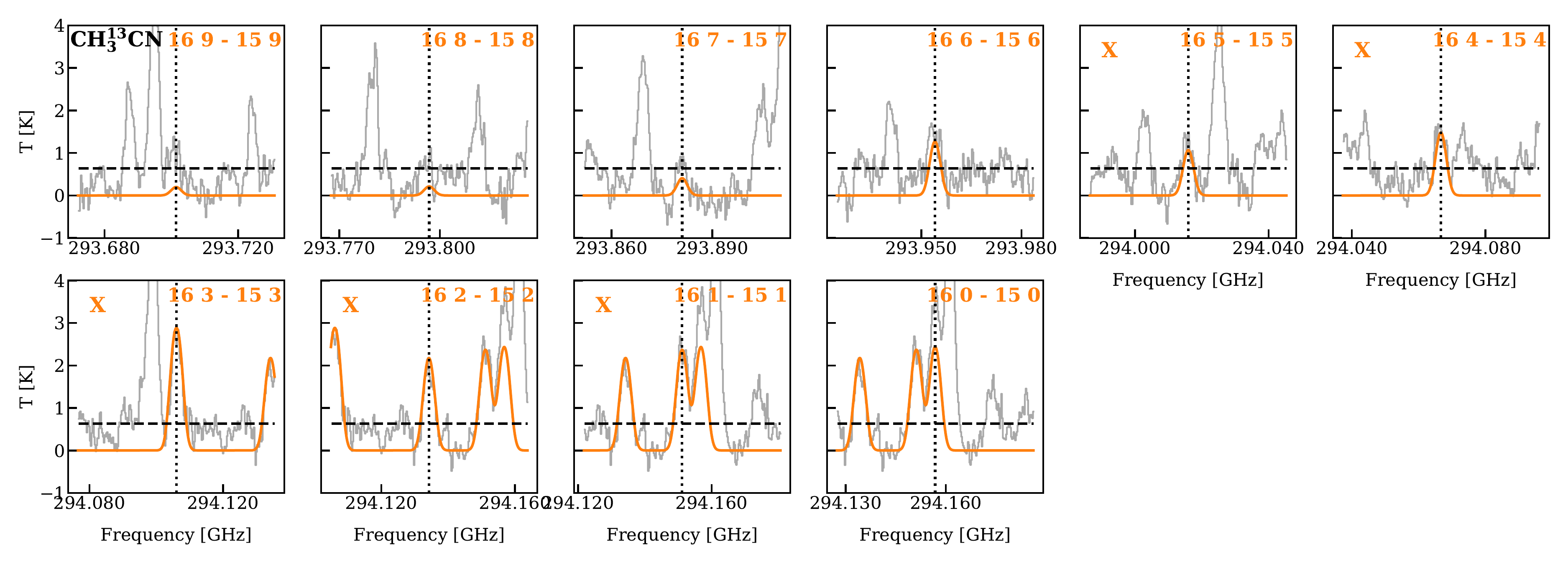}
    \caption{Same as Fig. \ref{fig:13CH3CN} but for CH$_3^{13}$CN.} 
    \label{fig:CH3C-13-N}
\end{figure*}

\begin{figure*}[h]
    \centering
    \includegraphics[width=\textwidth]{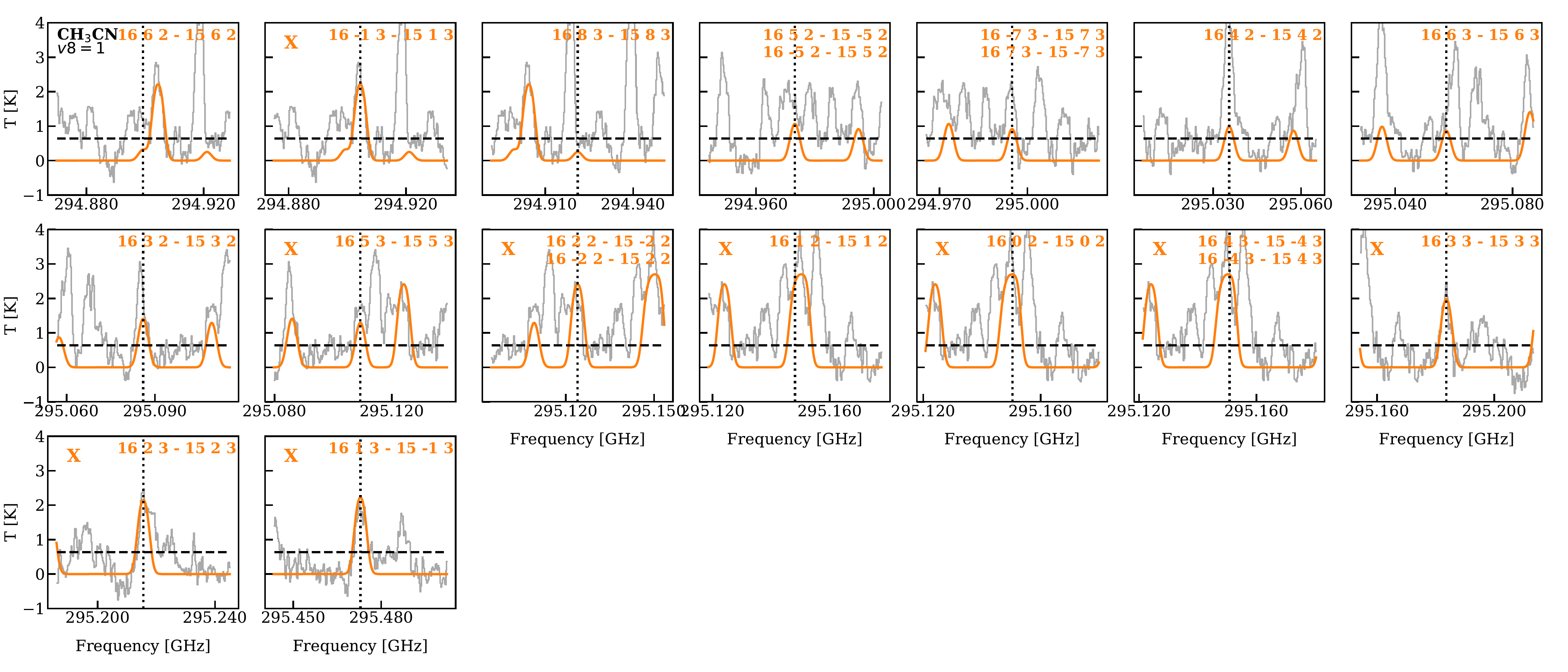}
    \caption{Same as Fig. \ref{fig:13CH3CN} but for CH$_3$CN $v_8$=1. Here, the lines with $E_{\rm up}< 1000$\,K and $A_{\rm ij} > 10^{-6}$\,s$^{-1}$ are shown.} 
    \label{fig:CH3CN_v8_1}
\end{figure*}

\begin{figure*}[h]
    \centering
    \includegraphics[width=19cm]{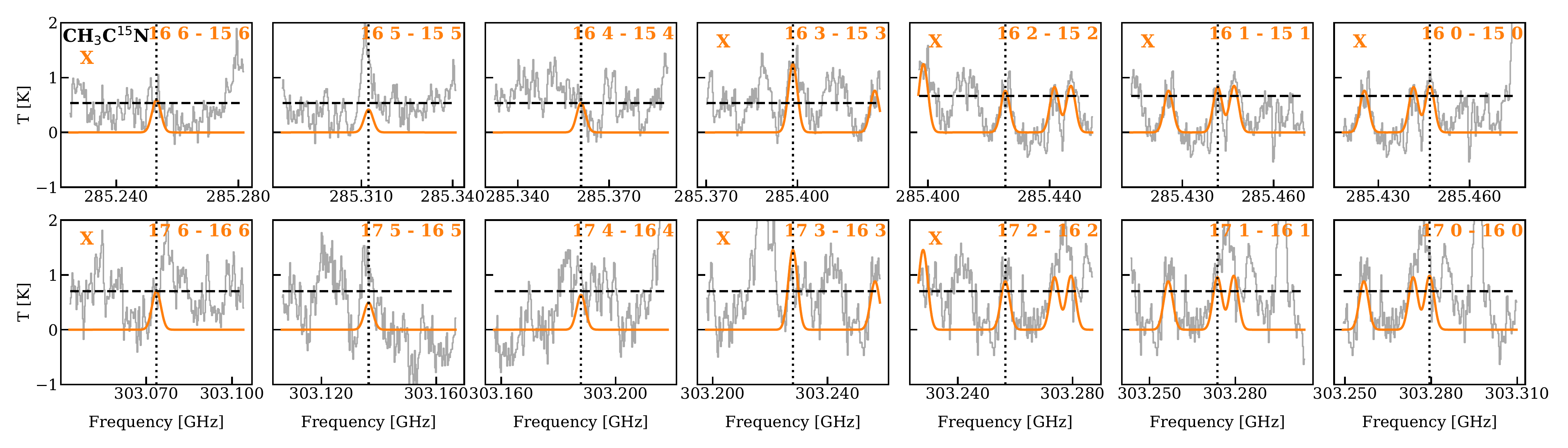}
    \caption{Same as Fig. \ref{fig:13CH3CN} but for CH$_3$C$^{15}$N. Here, only the lines with $E_{\rm up}< 400$\,K and $A_{\rm ij} > 10^{-3}$\,s$^{-1}$ are shown.} 
    \label{fig:CH3CN-15}
\end{figure*}

\begin{figure*}[h]
    \centering
    \includegraphics[width=19cm]{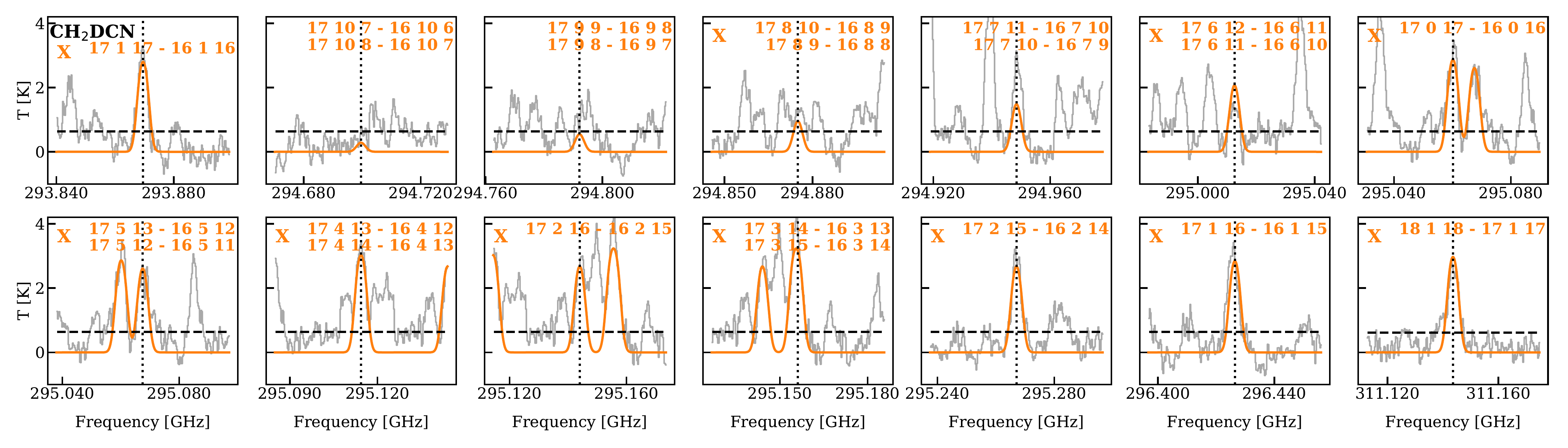}
    \caption{Same as Fig. \ref{fig:13CH3CN} but for CH$_2$DCN. Here, only the lines with $E_{\rm up}< 700$\,K and $A_{\rm ij} > 10^{-4}$\,s$^{-1}$ are shown.} 
    \label{fig:CH2DCN}
\end{figure*}

\begin{figure*}[h]
    \centering
    \includegraphics[width=19cm]{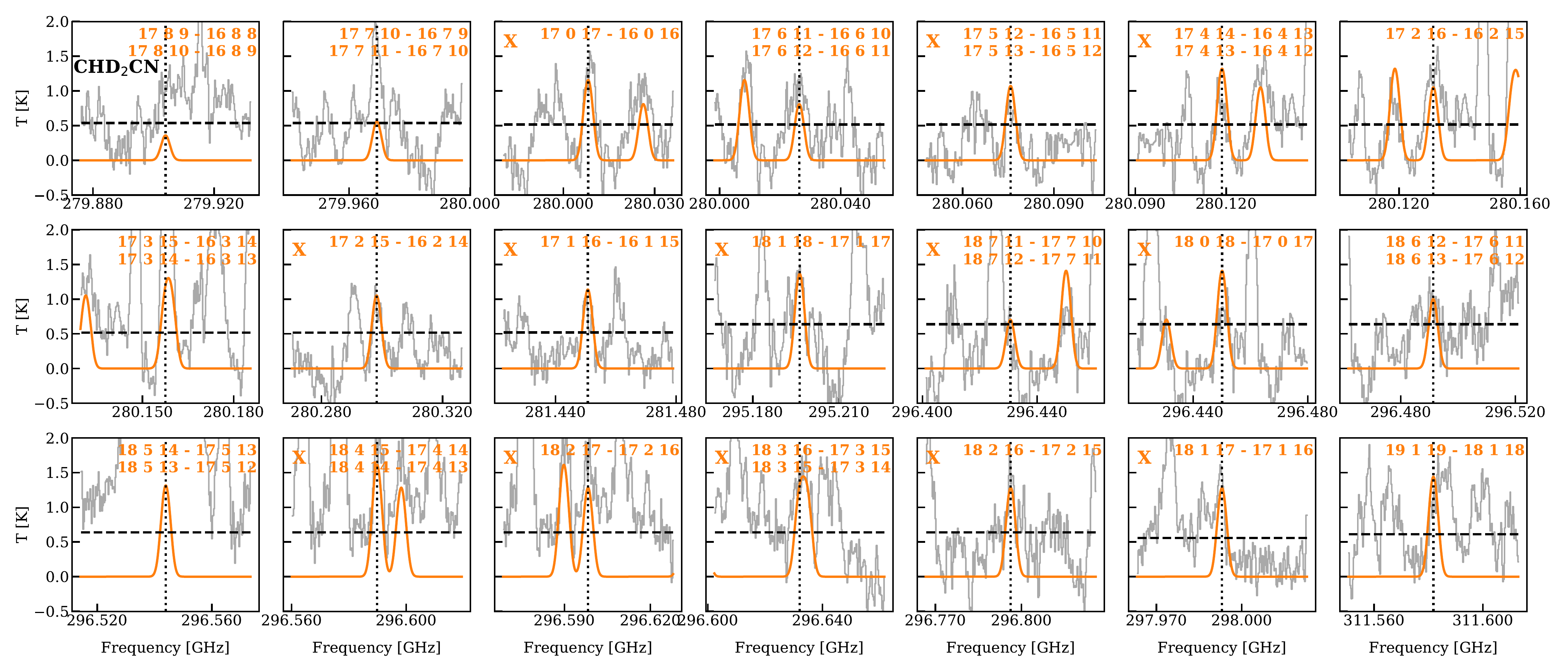}
    \caption{Same as Fig. \ref{fig:13CH3CN} but for CHD$_2$CN. Here, only the lines with $E_{\rm up}< 400$\,K and $A_{\rm ij} > 10^{-3}$\,s$^{-1}$ are shown.} 
    \label{fig:CHD2CN}
\end{figure*}

\begin{figure*}[h]
    \centering
    \includegraphics[width=19cm]{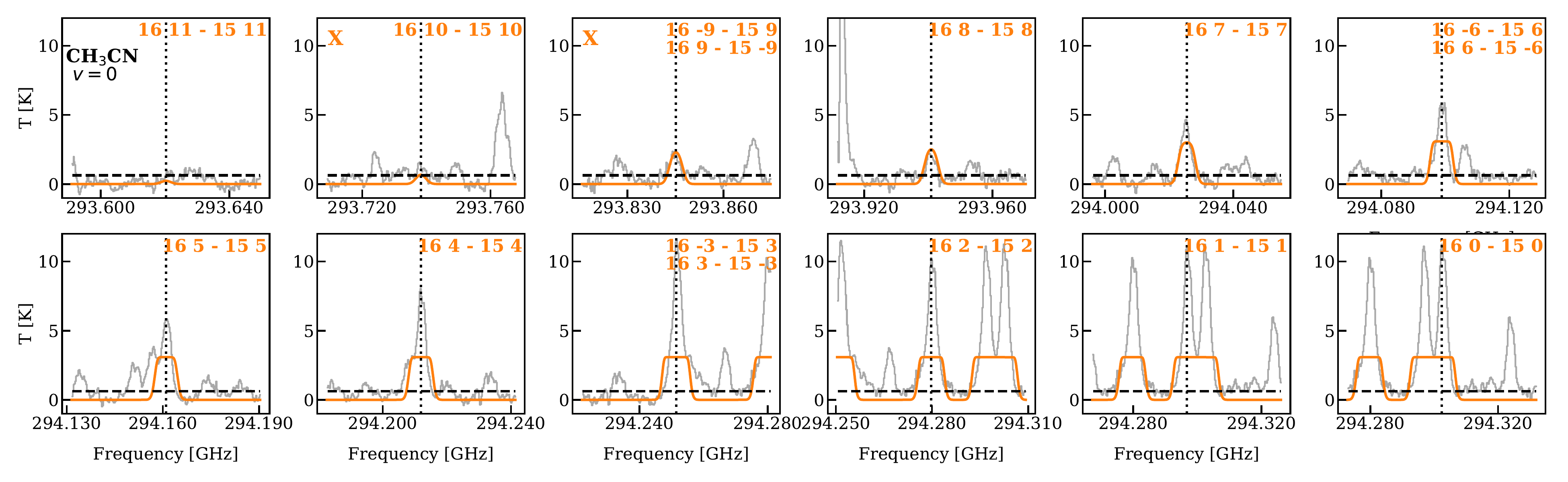}
    \caption{Same as Fig. \ref{fig:13CH3CN} but for CH$_3$CN, $v=0$. Only two potentially optically thin lines  were fitted while fixing the temperature to 150\,K.} 
    \label{fig:CH3CN}
\end{figure*}

Figures \ref{fig:CH3C-13-N}-\ref{fig:CHD2CN} present our CASSIS fits to the lines of various isotopologues. We note that all the lines that are covered in the data and are detected are presented in these figures. Figure \ref{fig:CH3CN} is similar but for CH$_3$CN and only showing the two potentially optically thin lines that were fitted. 

\end{appendix}

\end{document}